\documentstyle[12pt,epsf]{article}
\newcommand{\beq}{\begin{eqnarray}}   
\newcommand{\eeq}{\end{eqnarray}}
\newcommand{\ra}{\rightarrow}
\newcommand{\bq}{\begin{eqnarray}}   
\newcommand{\eq}{\end{eqnarray}}

\newcommand{\gsim}{\lower.7ex\hbox{
\;\stackrel{\textstyle>}{\sim}\;$}}
\newcommand{\lsim}{\lower.7ex\hbox{$
\;\stackrel{\textstyle<}{\sim}\;$}}

\begin{document}
\begin{titlepage}
\renewcommand{\thefootnote}{\fnsymbol{footnote}}

\begin{center} \Large
{\bf Theoretical Physics Institute}\\
{\bf University of Minnesota}
\end{center}
\begin{flushright}
OUTP-97-70-P\\
TPI-MINN-97/27-T\\
UMN-TH-1611-97\\
hep-th/9712046
\end{flushright}
\vspace*{1cm}

\begin{center}
{\Large \bf More on Supersymmetric Domain Walls, $N$ Counting   
and Glued  Potentials}

\vspace{0.8cm}

{\Large Ian I. Kogan,    Alex  Kovner}

\vspace{0.1cm}

{\it  Theoretical Physics, Oxford University. 1 Keble Road, Oxford 
OX13NP, 
UK}
\vspace{0.2cm}

 {\it and}

\vspace{0.5cm}

{\Large  Mikhail  Shifman} 

\vspace{0.1cm}
{\it  Theoretical Physics Institute, Univ. of Minnesota,
Minneapolis, MN 55455, USA}

\end{center}

\vspace*{.3cm}

\begin{abstract}

Various features of  domain walls in supersymmetric  gluodynamics
are discussed.  We give a simple field-theoretic interpretation of the 
phenomenon of strings ending on the walls  recently conjectured by 
Witten. An explanation of  this phenomenon in the framework of 
gauge field theory is outlined. The phenomenon is  argued to be  
particularly natural in supersymmetric theories which support
degenerate vacuum states with distinct physical properties.
The issue of existence (or non-existence) of the BPS saturated walls
in the theories with glued (super)potentials is addressed.  The 
amended  Veneziano--Yankielowicz effective  Lagrangian belongs to 
this class.  The physical origin of the cusp structure of the effective 
Lagrangian is revealed,  and the limitation it imposes on the 
calculability of the wall tension is explained.  Related problems are 
considered. In particular, it is shown that the so called discrete 
anomaly matching, when properly implemented, does not  rule out
the chirally symmetric phase of supersymmetric gluodynamics,
contrary to recent claims.

\end{abstract}

\vspace{1cm}

\end{titlepage}

\section{Introduction}

Recently there have been a renewed interest in the study of
theoretical aspects of $N=1$ supersymmetric (SUSY) gauge theories. 
In 
addition to
the calculation of exact effective potentials \cite{seiberg1}
and conjectured dualities between theories with distinct gauge
groups \cite{seiberg2}, it has been realized that  
in some supersymmetric theories there exists a class of dynamical 
objects
whose energy can be calculated exactly \cite{dvali1}. 
Those are the domain walls  interpolating  between 
discrete vacua which are typical for many SUSY gauge theories. The 
remarkable fact is that the energy
density (tension) of these walls is exactly calculable even in the 
strong coupling
regime. 

For 
supersymmetric gluodynamics,  the 
 theory of gluons and 
gluinos with no matter, 
the calculation of the energy density was carried out in Ref. 
\cite{dvali1}, in an indirect way.  The key ingredient is the
central extension of the $N=1$ superalgebra,
\beq
\{ Q^\dagger_{\dot\alpha}Q^\dagger_{\dot\beta}\}
=
\frac{N}{4\pi^2}\left(\vec\sigma\right)_{\dot\alpha\dot\beta}\int
 \, d^3 x \, \vec\nabla\left( \mbox{Tr}\, \lambda^2
\right) \, ,
\label{cext}
\eeq
where $Q^\dagger_{\dot\alpha}$ is the supercharge,
$\lambda$ is the gluino field, and  
$\left(\vec\sigma\right)_{\dot\alpha\dot\beta}
=\{ \sigma^3, -i , -\sigma^1\}_{\dot\alpha\dot\beta}
$ is a set of matrices converting the vectorial index
of the representation $(1,0)$ of the Lorentz group in the spinorial
indices.  The commutator 
(\ref{cext}) is given for $SU(N)$ gauge group;
the parameter $N$  stands for  the number of colors. The right-hand 
side of Eq. (\ref{cext}) is a reflection of ``geometric" anomalies
of SUSY gluodynamics (i.e. that in the trace of the energy-momentum 
tensor plus 
its supergeneralizations).

The integral over the full derivative on the right-hand side
is zero for all localized field configurations. It does not vanish, 
however, for the domain walls. Equation (\ref{cext}) implies that the 
tension  of the domain wall is
\beq
\varepsilon = \frac{N}{8\pi^2}
\left| \langle \mbox{Tr}\, \lambda^2\rangle_\infty
-\langle \mbox{Tr}\, \lambda^2\rangle_{-\infty}
\right| \, ,
\label{vacen}
\eeq
where the subscript $\pm\infty$ marks the values of the gluino 
condensate at spatial infinities (say, at $z\ra\pm\infty$ assuming 
that the domain wall lies in the $xy$ plane). The existence of the 
exact relation (\ref{vacen}) is a consequence of the fact that
the domain wall in the case at hand is a BPS-saturated configuration
preserving $1/2$ of the original supersymmetry.
A general discussion of BPS saturated domain walls was given in
\cite{chibisov},  while a particular  wall realization
in the framework of the amended \cite{kovner2} 
Veneziano--Yankielowicz 
effective Lagrangians \cite{veneziano}
were
studied in some detail in \cite{kovner1} and \cite{smilga}.

On the other hand, a theory  related to supersymmetric  
gluodynamics was analyzed 
recently from the point of
view of $D$-brane physics \cite{witten}. In this picture the domain 
walls
also appear naturally. Moreover they seem to have some rather 
surprising
properties. These properties are natural from the $D$-brane 
perspective but were considered unusual (even
paradoxical) from the field-theoretical point of
view \cite{witten}. One of such features is
an ``abnormal"  $N$ dependence of the wall tension. The wall energy
density of some  BPS saturated
walls scales as $N$, rather than $N^2$, a dependence one might 
expect {\em apriori} from 
 glueball solitons. 
The second surprise \cite{witten} is that the
confining QCD string emanating from the probe color  charges 
(quarks) on one side of the wall can
terminate on the wall,  without penetrating on 
the other side.

So far these features had no satisfactory 
explanation in  the field-theoretical framework.
One of our tasks is to understand how this works in field theory, at 
least at a qualitative level. We show that both aspects --
the $N$ dependence and termination of the flux tubes on the walls
-- are quite natural consequences of the peculiar gauge dynamics.
We suggest various toy models, which are simpler than SUSY 
gluodynamics but still carry essential features of the phenomena 
under discussion, to substantiate our qualitative observations. 
For instance, an Abelian model is presented
where probe  fractional charges generate induced ``mirror" 
fractional charges on the wall in spite of the fact that
at the fundamental level the model contains no fractionally charged 
fields. 

Then we turn to the Veneziano--Yankielowicz effective Lagrangians.
Previously they were exploited as a framework for 
quantitative analysis of the domain walls \cite{kovner1}. A BPS wall 
interpolating  between a  chirally symmetric vacuum \cite{kovner2} 
and one of the conventional vacua with $\langle\lambda\lambda
\rangle \neq 0 $ was explicitly built.
However, building a wall interpolating between two
neighboring chirally asymmetric vacua turned out to be a much 
harder endeavor. The task had not been solved in Ref. 
\cite{kovner1}. Moreover, it was argued later \cite{smilga}
that such walls do not exist 
within the framework of the Veneziano--Yankielowicz effective 
Lagrangians.

A crucial feature of such a  Lagrangian emerging in SUSY 
gluodynamics is discontinuity of the superpotential ${\cal W}$.
The realization compatible with all symmetries of the underlying 
theory requires, with necessity,  a ``glued" potential 
\cite{kovner2},
with $N$ distinct sectors and matching lines along the boundaries of 
the sectors.  We explain the physical nature of this phenomenon.
The sector pattern, with cusps, reflects a restructuring of heavy 
degrees of freedom (which were integrated out) in the process of an 
adiabatic variation  of the light degrees of freedom. A level crossing 
takes place in the heavy sector of the theory. 
Precisely for this reason, one {\em cannot} construct the domain wall
from the effective Lagrangian if the wall crosses the cusp.
The presence of
the cusps prevents one from being able to use this potential 
for calculating a wall 
profile if the field configuration along the wall crosses the cusp 
somewhere
in space. 
In fact, if one naively tries to do this 
in the presence of the cusp an apparent 
paradox arises --  the
wall in the effective theory seems to have a lower energy density 
than the BPS bound on this quantity in
the original theory. The
missing energy density is contributed 
by the excitation of the heavy modes which are
necessarily excited when the light fields take values in the vicinity of 
the cusp.

The statement is thoroughly illustrated by two toy models.
The phenomenon is quite general and may be considered in
supersymmetric as well as non-supersymmetric context. 

The  chirally symmetric vacuum, $\langle\lambda\lambda
\rangle = 0 $, is inherent to  the  Veneziano--Yankielowicz  
Lagrangian.
Recently it was claimed \cite{CH} that a discrete anomaly matching
rules out the existence of such phase, at least its most 
straightforward realization. We make a digression to show that
the  claim is due to an inconsistent treatment of the discrete 
anomaly matching. In SUSY gluodynamics and similar theories
the discrete 
anomaly matching imposes  no constraints on the spectrum. The  
only information one gets are rather 
mild constrains  on certain  amplitudes following from the classical 
symmetries that become anomalous at the quantum level
(see e.g. \cite{NSVZ}).
The existence of a discrete anomaly-free subgroup
adds no new information. 
 
The organization of the paper is as follows. Section 2 is devoted
to general aspects of the BPS walls in SUSY gluodynamics. 
We analyze, qualitatively,   the $N$ dependence of the 
wall tension and visualize the
phenomenon of strings ending on the 
domain walls. An analogy between the 
walls in SUSY gluodynamics  
 and the axion  domain walls in gauge theories with monopoles
is worked out.

In Sect. 3 we turn to a more quantitative discussion based on the
amended 
Veneziano--Yankielowicz effective action. The issue of glued 
potentials is studied here. 
In Sect. 3.3 we consider an explicit example of a supersymmetric 
theory 
which upon
integrating out heavy 
modes generates an effective potential with cusps. In this
model we calculate explicitly 
the cusp contribution to the wall tension and  show how
the apparent contradiction with the BPS bound is resolved.
In Sect. 3.4 aspects of the general theory of glued (super)potentials
are presented. 

In Sect. 4 we discuss other domain walls in gauge theories
obtained as a Kaluza-Klein reduction on topologically non-trivial 
space-time manifolds. The specific example considered refers to
$R_4\times S_1$. An
 unconventional  $N$ dependence of the wall tension arises which 
 may be related to $D$-branes.

Section 5 is devoted to the issue of the discrete anomaly
matching and the chirally symmetric vacuum of SUSY gluodynamics.

Finally, Sect. 6 contains a brief summary and  discussion of our 
results.
 
\section{$N$ Dependence, Flux Tubes Ending on the Walls and All 
That}

\subsection{SUSY Gluodynamics}

We consider the supersymmetric 
generalization
of pure gluodynamics -- i.e. the theory of gluons and gluinos. At the 
fundamental level the 
Lagrangian of the model  has the form 
\cite{FZ}
\beq
{\cal L} =  -\frac{1}{4g_0^2} 
G_{\mu\nu}^aG_{\mu\nu}^a + {\vartheta\over 
32\pi^2}G_{\mu\nu}^a\tilde G_{\mu\nu}^a
+\frac{1}{g_0^2}\left[
i\lambda^{a \alpha} 
D_{\alpha\dot\beta}\bar\lambda^{a\dot\beta}
\right] \, 
,
\label{SUSYML}
\eeq
where the spinorial notation is used. In the superfield language
the Lagrangian can be written as
\beq
{\cal L} =  \frac{1}{4g^2} \mbox{Tr} \int d^2\theta W^2 + \, 
\mbox{H.c.}
\, ,
\label{SFYML}
\eeq
where
$$
\frac{1}{g^2} = \frac{1}{g_0^2} -\frac{i\vartheta}{8\pi^2}\, .
$$
Here  $\vartheta$ denotes the vacuum angle.  Our conventions 
regarding the superfield formalism 
are summarized e.g. in the recent review
\cite{Shif2}. We will limit ourselves to the $SU(N)$ gauge group
(the generators of the group $T^a$ are in the fundamental 
representation, so that $\mbox{Tr}(T^aT^b) = (1/2)\delta^{ab}$).

$SU(N)$ supersymmetric gluodynamics 
has a discrete symmetry,  $Z_{2N}$,   a (non-anomalous) 
remnant of the 
anomalous axial 
symmetry generated by the phase rotations of the gluino field. The 
gluino condensate $\langle\lambda \lambda\rangle$ 
is the order parameter of this symmetry.  
The discrete chiral 
symmetry may or may not be spontaneously broken \cite{kovner2}.
Therefore, there exists a set of distinct vacua labeled by the
value of the gluino condensate. In the phase with the broken chiral 
symmetry Tr$\langle\lambda \lambda\rangle
=\Lambda^3 \exp (2\pi i k /N)$ where $k=0,1,2 ,..., N-1$
(for $\vartheta=0$). In the chirally 
symmetric
phase  Tr$\langle\lambda \lambda\rangle = 0$.
The field configurations interpolating 
between different values of $\langle\lambda \lambda\rangle$ 
at spatial infinities are topologically stable domain walls.
Although the theory is  in the strong coupling regime  one can 
derive 
an exact 
lower bound on the
surface energy density (tension) for such a wall \cite{dvali1}
\beq
\varepsilon\ge \frac{N}{8\pi^2}
\left| \langle \mbox{Tr}\, \lambda^2\rangle_\infty
-\langle \mbox{Tr}\, \lambda^2\rangle_{-\infty}
\right|\, .
\label{vacen1}
\eeq
In our normalization the condensate $\langle \mbox{Tr}\, 
\lambda^2\rangle$ scales as $N$
in the large $N$ limit. 

One may  consider 
two types of walls. The wall of
type I connects a vacuum 
with the spontaneously broken chiral symmetry with the symmetric
vacuum ($\langle \mbox{Tr}\, \lambda^2\rangle=0$). 
For such a wall the BPS bound for the tension is
$$
\varepsilon\sim O(N^2)\, .
$$ 
The walls of type II
connect two adjacent (or close) 
chirally asymmetric vacua, e.g. $k=0$ and $k=1$ (or $k=0$ and $k=2$, 
etc.). Even though for each of these vacua
the order parameter is of order $N$, 
the difference between the order parameters is
$O(1)$. The BPS bound for the tension, therefore,
 is 
$$
\varepsilon\sim O(N)\, .
$$

Let us assume for the moment 
that the BPS-saturated walls of type II do indeed exist in 
SUSY gluodynamics. Although we cannot prove this at present,
there are no visible reasons forbidding them \footnote{
Even if for some reasons we do not understand at present the type II 
walls would turn out to be not BPS-saturated, it is natural to expect 
that their tension is of order of the BPS bound (\ref{vacen1}).
}.

The question then arises as to how one can  understand the large 
$N$ scaling of the wall energy density
from the point
of view of the effective field theory which 
describes  dynamics of the low lying physical
states,  mesons and glueballs and their superpartners.

At large $N$ the mesons and the glueballs 
should have masses of order 1, trilinear couplings of
order $1/N$, and so on \cite{HW}. This is conveniently 
encoded in an effective Lagrangian
of the form
\begin{equation}
{\cal L}=N^2F[M_i, \partial M_i]\, ,
\label{glueb}
\end{equation}
where $\{M_i\} $ is a set of fields
representing all relevant degrees of freedom, mesons and glueballs.
The value of the functional $F$ 
itself and all its derivatives at the minima should be
independent of $N$. This would ensure the proper $N$
dependence of the masses and coupling constants. 
Now, suppose we have a solution of classical equations of motion 
$M_{\rm wall}$
which describes
a wall
configuration interpolating between two distinct minima.
Since $N^2$ is an overall factor in Eq. (\ref{glueb}), 
at first  sight one may expect that $F[M_{\rm wall}]=O(1)$, 
and, therefore,   the wall tension
$\varepsilon=O(N^2)$. Such a situation is standard
in the soliton physics. 
This is perfectly in 
line with what we get for  the type I walls,  but is in apparent 
contradiction with the BPS energy density
of the type II wall.

Can one avoid this conclusion?
The answer is yes. Consider a function $F$ which is nonanalytic so that
although all its derivatives are $O(1)$ at the minimum $M_*$, at some
finite distance $M=M_*+\delta M$  (with $\delta M=O(1/N)$) the
derivatives become large ($O(N)$, for example). Then, if the wall solution
$M_{\rm wall}$ passes through this region of the field space,
the standard counting leading to $\varepsilon=O(N^2)$
does not work.
An extreme example of such a situation arises
if the effective Lagrangian has a structure of $N$ distinct sectors 
in the space of fields,  and  no single analytical function $F$ exist. 
The $N$ sector structure then is an implicit source of $N$ dependence.
Such a potential is singular (or rather has a singular first
derivative) 
along the boundaries of the sectors, being glued 
from different pieces 
along the boundaries. As we will discuss later, this situation can
arise due to the fact that the state that
was the ground state in one sector,
becomes an excited state in another, and vice versa.
At the boundary  there are 
 degenerate states, and  the level crossing occurs. 
Due to the cusps at the boundary the  naive estimate of the tension
presented above does not work in the case of 
the glued potential. As we will see, the effective potential in SUSY
gluodynamics has precisely this $N$--sector structure.

Of course, in the full theory everything is smooth. One can ensure the 
smoothness of an  effective Lagrangian by including more fields in it. Those
extra fields will not correspond to low energy excitations in the vicinity
of the minima (and therefore will be unimportant for local properties
like Green's functions), but will be essential for smoothing the
singularity at the cusps. In the example just discussed one would have
to  include in the game at least $N$ fields.
That is how $N$ enters the effective potential as a hidden parameter
(besides the overall $N^2$ factor in Eq. (\ref{glueb})).
Then the typical value of  relevant fields $M_i$ inside the wall 
solution can be $O(N^{-1})$, each field contributes to $F$ at the level
$O(N^{-2})$, but there are $N$ relevant fields, 
and the value of $F\sim 1/N$. Note that the wall width is
$L\sim N^0$. Then the volume energy density inside the wall
$E \sim N$ and, correspondingly, the wall tension
$\varepsilon \sim N$.

The fact that the volume energy $E $ is
$O(N)$  inside the BPS wall 
connecting two neighboring vacua, say with Tr$\langle\lambda 
\lambda\rangle
=\Lambda^3 \exp (2\pi i k /N)$ where $k=0$ and 1, is seen in  the 
microscopic theory (\ref{SUSYML})
{\em per se}.  Indeed, for the BPS wall
$$
G_{\mu\nu}^a G_{\mu\nu}^a \sim \partial_z \lambda^2
\sim \Delta \lambda^2 /L \, .
$$
 Since the volume 
energy density $E\sim N \, G_{\mu\nu}^a G_{\mu\nu}^a$,
and $\Delta \lambda^2$ in the neighboring chirally asymmetric 
vacua is $O(1)$, we conclude that $E$ scales as $N$.

How do we learn about the $N$-sector structure of the
effective Lagrangian emerging in SUSY gluodynamics?
If the gauge group is  $SU(N)$, the theory  has a
discrete $Z_{2N}$ chiral symmetry, which is 
spontaneously broken down to $Z_2$ in some
of the vacua. This means that the 
effective Lagrangian for mesons/glueballs must have at least $N$
degenerate minima which differ from each other only in 
the value of the phase $\phi$
of the order parameter Tr$\lambda^2$ 
(the latter is an interpolating field for  one of the lighter mesons). 
The minima lie at $\phi = 0, 2\pi / N, 4\pi /N$ and so on.
Then,  clearly,  we must have a much more rapid variation of $F$ 
as a function of $\phi$, than one would naively infer.
Naively, since there is no explicit $N$ dependence in $F$,
one would say that if the first minimum lies at  $\phi = 0$, the 
second one should be at $\phi \sim 1$. 
The only way out is either to have an $N$-sector structure (within 
the construction that includes only fixed, $N$  independent number of
fields in the effective Lagrangian), or to build a Lagrangian on a minimal 
set of
$N$ fields. In both cases a hidden parameter $N$ appears.
It does not affect the value of the derivatives of $F$ at the minima,
which are all of order 1. 

In Sect. 3 we will consider the 
amended Veneziano--Yankielowicz effective
Lagrangian and will see that it indeed has the required  structure.
In a somewhat simplified picture,  we can understand how it appears
 by considering the dependence
of the vacuum energy in the Yang-Mills theory on the vacuum angle
$\vartheta$. 
As is well
known form the consideration of the Ward identities 
\cite{wardident}, 
 this dependence has
a ``wrong" $\vartheta$ periodicity. That is, naively the energy is
periodic in $\vartheta$ with the period $2\pi N$ rather than $2\pi$,
\begin{equation}
E_{\rm vac} \propto N^2\left[ \cos{\vartheta\over N}- 1\right] \, .
\end{equation}
The correct 
periodicity of the physical quantities is restored 
in the following way. The ground state at $\vartheta <\pi$,
becomes an excited state at $\vartheta >\pi$.
At $\vartheta =\pi$ there are 
two degenerate states. At this point due to the level crossing the
vacuum energy has a
cusp so that
\begin{equation}
E_{\rm vac} \propto N^2\left[ \left( \cos{{\vartheta-2\pi k}\over N}
\right) - 1\right] \, , \ \ \ \ 
(2k-1)\pi<\theta<(2k+1)\pi \, .
\label{thetadep}
\end{equation}
Now, the interaction of the phase $\phi$ with the gluonic
degrees of freedom is the same as of the rescaled angle  
$\vartheta/N$.
The effective potential for $\phi$ is, therefore, roughly
\begin{equation}
E_{\rm vac} \propto N^2\left[ -1 + \cos\left( \phi-{2\pi k\over 
N}\right) 
\right]\, , \ \ \ \ \ 
{(2k-1)\pi\over N}<\phi<{(2k+1)\pi\over N}\, .
\label{phidep}
\end{equation}
This potential indeed has the form of Eq. (\ref{glueb}). Moreover, the 
derivatives of the function $F$ at all the minima are 
$O(1)$. Nevertheless, it has $N$ 
minima at $\phi={2\pi k / N}$.  The value of $F$ on interpolating 
trajectories varies from zero at the minima to $O(N^{-2})$ at the cusp.
A naive estimate of the wall tension would therefore give a value
much smaller \footnote{In fact, a naive (and wrong)
estimate would give
$O(N^{0})$. Actually, a much larger contribution, $O(N^{1})$, resides
in the cusp, see Sect. 3.} than $O(N^2)$.

Our next remark concerns a 
field-theoretical understanding of a confining string which
ends on a domain wall. 
A simple example of such a situation is the wall that separates the
confining 
phase in a gauge
theory from a nonconfining one. 
The type I wall in SUSY gluodynamics is precisely of this
kind. Recall that the chirally symmetric 
vacuum at $\langle \lambda^2\rangle =0$ sustains massless excitations.
It was argued in \cite{kovner2} that this 
phase is in fact conformally invariant and can
be thought of as a kind of a non-Abelian Coulomb phase.
Now consider a wall that separates a confining phase from a 
Coulomb phase. A probe charge
placed in the confining phase 
is a source of the electric flux which travels in a flux
tube --
the confining string. On the other side of the wall, 
however, it is energetically favorable
for the flux to spread out into a Coulomb field. 
So an observer in the Coulomb phase will
not see a string but, rather, a point charge (in fact, twice as big in 
magnitude as the
original probe  charge, since the electric 
flux will spread in half the space) sitting on
the domain wall. 
One is not used to thinking about a phase boundary between the 
Coulomb 
and confining phases,
since usually the two 
are not degenerate in energy. It is the peculiar feature of
supersymmetric theories 
that two physically completely distinct phases are degenerate.

It is very easy to find  an easy-to-handle
example of a domain wall separating the confining and the Abelian 
Coulomb 
phases by introducing some extra fields
in SUSY gluodynamics. 
Start from the Lagrangian (\ref{SFYML}), and add one chiral matter 
superfield $\Phi$ in the adjoint representation of the gauge group, 
with the  superpotential
\beq
{\cal W} = m\mbox{Tr}\Phi^2 + M^{-1} (\mbox{Tr}\Phi^2)^2
\label{potents}
\eeq
(the gauge group  $SU(2)$ is assumed). The second term in the
superpotential is non-renormalizable. One can think of it as
a result of integrating out some heavy degrees of freedom,
so that at a large scale $M$ we return back to a renormalizable theory.
It is assumed that $m\ll\Lambda$,  but
$m M \gg\Lambda^2$.

If $ M^{-1} = 0$  the theory we deal with is nothing but a softly 
broken $N=2$ model
studied by Seiberg and Witten \cite{SEIWIT}. In the Seiberg-Witten 
vacua $\mbox{Tr}\Phi^2\sim \Lambda^2$,  monopole condensation 
takes place,
and due to the dual Meissner effect the probe electric charges 
placed in one of these vacua will form  flux tubes. The presence of a 
very weak additional interaction  $M^{-1}( \mbox{Tr}\Phi^2)^2$ 
not considered in Ref. \cite{SEIWIT} does not affect the picture
obtained there, since this term can be viewed as an arbitrary  small 
perturbation if the theory resides in one of the Seiberg-Witten vacua.

However, at large values of  $\mbox{Tr}\Phi^2$ the term $M^{-1}
(\mbox{Tr}\Phi^2)^2$ leads to a drastic restructuring  -- no matter how 
small $M^{-1}$ is there appears a new vacuum state at
$\mbox{Tr}\Phi^2 \sim m M $. In this vacuum the gauge 
symmetry is  broken down to $U(1)$ by a very large vacuum 
expectation value of the $\Phi$ field, the monopoles are 
very heavy, and the theory is obviously in the weakly coupled 
Coulomb phase. Supersymmetry guarantees that the vacuum energy 
densities in both phases vanish: the two phases are degenerate.
Under the circumstances a domain wall separating  the weakly 
coupled Coulomb phase and the strongly coupled confining phase 
(one of the Seiberg-Witten vacua) must exist, with the wall tension 
$\sim m^2 M $. If the confining phase is to
the left of the wall, and we put there a probe electric charge,
a flux tube going towards the wall develops; the 
chromoelectric flux is clearly diffused to the right of the wall. 

Another example of a wall that serves 
as a sink of the chromoelectric flux is a situation when  the Coulomb 
phase
exists not in half the space (say, to the left of the wall) but 
only inside the wall. For example if one considers a wall
in SUSY gluodynamics that separates the phases 
Tr$\lambda^2=\Lambda^3$ and
Tr$\lambda^2=-\Lambda^3$, 
it is very likely
that the order parameter $\lambda^2$ will vanish inside the wall. 
Both
phases 
then are
confining but the wall itself is ``made" of the 
Coulomb vacuum. In such a case the flux
tube that originates in one of the phases will not penetrate into 
another but, most
likely, the flux will spread out in transverse directions inside the 
wall 
in a two-dimensional Coulomb field. Energetically
this is preferable, since the energy
of the two-dimensional Coulomb field 
depends only logarithmically on the size of
the system, while the energy of the 
string that penetrates into the other phase is linear.

Note that in this latter scenario the degeneracy between the Coulomb
and confining phases is 
not necessary. The picture can be dynamically realized
both in the supersymmetric and nonsupersymmetric
contexts. An illustrative nonsupersymmetric model,
where inside the wall the theory is in the Abelian Coulomb phase
while outside it is in the confining phase, was presented in
Ref. \cite{dvali1}.
It would be instructive to exploit this model for a more quantitative
analysis
of a flux tube coming from the confining phase and diffusing itself
inside the wall (i.e. in the Coulomb phase). A (semi)quantitative
analysis
seems possible since at least inside the wall the theory \cite{dvali1}
is in the weak coupling regime.

Finally, 
if in the previous examples we consider the Higgs phase instead of 
the Coulomb
phase (either to the left of the wall, or inside it), the 
chromoelectric flux will still disappear in the wall. In this case it is 
even 
more
trivial, 
since the chromoelectric flux is not conserved in the Higgs phase, and 
it will 
be
 screened 
by the Higgs phase vacuum either on the left  side or inside the 
wall.
We would like 
to argue that the type II wall,   considered in Ref. \cite{witten},
 is, in fact,  an example of
this kind, in a certain sense.
Indeed, consider the 
  type II wall. Let us say, on the left there lies  a phase 
with the 
condensate of
monopoles, while 
on the right  with the condensate of dyons
\footnote{Both, the monopoles and dyons in SUSY gluodynamics are 
to be understood in the same sense as those in  the 't Hooft 
construction \cite{hooft}.}. Let us imagine
a probe electric charge to the left of the wall. Since the dyons 
are
electrically charged, their condensate acts like a Higgs vacuum, in the 
sense
that it can be 
easily polarized to completely screen the electric flux that might 
enter
the dyon condensate through the wall. Of
course, since the dyons are
also magnetically charged, any such polarization of their condensate
will lead to appearance of net magnetic charge to the right of the 
wall.  
However, the magnetic flux tube emanating from this induced 
magnetic charge can 
be directed towards the 
domain wall. In that case it will be screened on the other side of
the wall by polarization of the condensate of the monopoles.
In other words 
it is plausible that the dyonic condensate to the right of the wall will
be polarized to screen the electric charge while the monopole 
condensate to the
left of the 
wall will be polarized to compensate for the excess of the induced 
magnetic
charge. As a result the confining electric string will terminate on the 
wall.

Note, that this picture is somewhat 
different from the one advocated in Ref. \cite{witten}, where it was
suggested that a bound state of a monopole and a dyon appears on 
the wall. 
It is difficult to talk about monopoles and dyons forming a bound
state, since they do not exist as free particles on either side of the
wall, because  both vacua have
nonvanishing condensates.

\subsection{Toy model -- axion wall}
The type II wall can be thought of as
carrying
the  quark quantum numbers
 in the  presence of a   QCD string, in the sense that
it can screen a fundamental charge or other charges which are 
nontrivially transformed
 under the  center $Z_{N}$, in the  theory where all 
dynamical
fields  are  invariant with respect to $Z_{N}$. 
Surprising as it is,  one can trace the very same  phenomenon in 
simple Abelian models. Although the parallel is not perfect,
a simple Abelian example may be useful for deeper understanding of this
general phenomenon.

The problem we keep in mind is the
 axion wall in the  presence of a
 monopole \cite{kogan}. For simplicity we will consider the $SU(2)$ 
case 
(the Georgi--Glashow model). The   $SU(2)$ symmetry is 
spontaneously broken by the vacuum expectation value of a Higgs 
field down to $U(1)$ giving rise to the 't Hooft-Polyakov monopoles
\cite{THP}. 
After the breaking,  the fields in the  adjoint representation have the 
$U(1)$  charges $\pm 1$, while those in the  
fundamental representation have  charges $\pm 1/2$.  
 
 Let us recall some facts about  the  monopoles in 
the presence
 of the $\vartheta$ term. The Lagrangian  of the Georgi--Glashow 
model  is 
\bq
{\cal L} = -\frac{1}{4e^{2}}F_{a}^{\mu\nu}F^{a}_{\mu\nu}
 + \frac{\vartheta}{32\pi^{2}}\tilde{F}_{a}^{\mu\nu}F^{a}_{\mu\nu}
 + L_{H}(\Phi)\, ,
\label{lagr1}
\eq
where $e$ is the gauge coupling constant and 
the last term is the scalar Lagrangian for the  Higgs field
 $\Phi^{a}$ in the adjoint representation
 of the  $SU(2)$ group. 

 It was shown by Witten
\cite{witten2} that  if $\vartheta\neq 0$,  a monopole becomes 
a dyon with the electric charge   
\bq
q = \frac{\vartheta e^2}{8\pi^2} \mu + ne\, ,
\label{witten}
\eq
where $\mu$ is the magnetic charge,
\bq
\mu = \frac{4\pi}{e}\, .
\eq
When $\vartheta$ changes from $0$ to $2\pi$  one gets $n 
\rightarrow n + 1$. 

In the theory (\ref{lagr1}) $\vartheta$ is constant, given once and 
forever.  However, if the axions are added in the theory,
then, effectively,  $\vartheta$ is substituted by
the axion field which can vary in space-time. The axion field can be
introduced through a phantom-axion construction \cite{SVZ}, i.e.
we add an $SU(2)$-singlet Higgs field coupled to a doublet
quark field. In the limit when the expectation value of the
$SU(2)$-singlet Higgs field tends to infinity, the quark becomes 
infinitely massive and disappears from the spectrum, and so does
the modulus of the singlet Higgs field. Its phase
becomes an axion field $a(x)$.

The axion  Lagrangian is 
\bq
{\cal L}_a =\frac{ f_{a}^{2}}{2}
\partial_{\mu}a \partial_{\mu}a -
K^{2}\left[ 1-\cos (a -\vartheta )\right] \, , 
\label{Ltheta}
\eq
where the  parameter $K^{2}$ is connected with the  vacuum 
susceptibility, it is exponentially small in the model at hand, $\sim 
\exp (-8\pi^2 /e^2)$. Moreover, $f_a$ is
(a very large)  expectation value of the singlet Higgs. In this limit the 
only other axion interaction to be taken into account is
its coupling to $F\tilde F$.  The $\vartheta$ term in Eq. (\ref{lagr1})
becomes
\bq
{\cal L}'_\vartheta = 
\frac{\vartheta 
-a(x)}{32\pi^{2}}\, \tilde{F}_{a}^{\mu\nu}F^{a}_{\mu\nu} \, .
\label{extral}
\eq
One can now set $\vartheta = 0$ (and we will do this hereafter). 

The vacua at $a = 0$ and $a = 2\pi$ are physically identical.
Correspondingly, 
  axion domain walls exist in this system  \cite{axiondomainwall}
interpolating between $a = 0$ and $a = 2\pi$. Assume a wall  lies in 
the 
$xy$ plane, so that the axion profile depends only on 
 $z$. The wall solution centered at  
origin is
\bq
a (z) = 2\pi - 2\arccos \tanh ( mz)
\eq
where $m$ is the axion mass,
$m = K/f_a$, 
 and the width  of the wall  is of  order $m^{-1}$. 
  Of course, it is assumed that $m\ll M$ where $M$ is the monopole 
mass. 

Start from  the monopole with electric charge zero
to the left of the wall, and let it adiabatically propagate  
through
 the wall. Effectively, $\vartheta$ adiabatically changes
from 0 to $2 \pi$. To the right of the wall the monopole becomes a 
dyon with electric charge 1.  Inside the wall, the electric charge of 
the monopole gradually increases. 

 Thus,  one gets an apparent  
 nonconservation of  the  electric charge of the monopole. However,
since $U(1)$ is unbroken , the total electric charge   must be 
conserved. The question is where is the missing electric charge. 

 As  was shown by Sikivie \cite{sikivie}, the monopoles actually  
induce  
electric charges $\pm 1/2$ on the wall. 
 When the monopole is far to the left of the wall, it is 
  neutral, but the  charge induced on the wall is  $+1/2$. 
When the monopole is far to the right of the wall, 
 the induced charge will be $-1/2$, so the total charge is conserved, 
 $1/2 = 1 - 1/2$.

To see that this is indeed the case we observe that the extra term
(\ref{extral}) in the Lagrangian is immediately translated into an 
additional piece in the definition of the electromagnetic current
\bq
j_\mu^{\rm axion} = \frac{\delta {\cal L}'}{\delta A_\mu } = 
\frac{1}{8\pi^2}\, \partial_\nu \left(
a \tilde{F}_{\mu\nu}^3\right)
\label{excur}
\eq
where it is assumed that the expectation value of the triplet Higgs 
field is aligned along the third direction, so that ${F}_{\mu\nu}^3$
is the photon field strength tensor. Note that $j_\mu^{\rm axion}$
is automatically conserved, $\partial^{\mu}j_\mu^{\rm axion} = 0$.
The corresponding contribution in the electric charge consists
of two parts,
\bq
Q= \int d^3 x j_0^{\rm axion} = \frac{1}{8\pi^2}\, 
\left(\tilde{F}_{0\nu}^3 \partial_\nu a + a\partial_\nu 
\tilde{F}_{0\nu}^3\right)\, .
\label{twoterm}
\eq
Let us assume that the distance between the wall center $R$
and the monopole is much larger than $m^{-1}$. For such distant
monopoles the physical meaning of each term in Eq. (\ref{twoterm})
is transparent. The second term vanishes everywhere except
the point where the monopole sits. Thus, it gives the electric charge 
of the monopole/dyon. If the latter sits to the left of the wall,
where $a=0$, the monopole electric charge vanishes. To the right of 
the wall $a= 2\pi $, and
$$
\Delta Q_{\rm monopole} = \frac{e}{4\pi }\int  d^3 x 
\vec\nabla \vec B_{\rm monopole}  = \frac{e\mu }{4\pi } = 1\, .
$$
The first term in Eq. (\ref{twoterm}) is obviously saturated inside 
the wall; it describes the electric charge induced on the axion wall
in the presence of a distant monopole. The induced charge is equal to 
the flux of the monopole magnetic field through the plane of the wall
times $\Delta a  /(8\pi^2 )$. Since this flux is 1/2 of the flux through 
the large sphere ($= e\mu/2 = 2\pi$), the induced charge on the wall 
is obviously equal to
$\pm 1/2$ depending on whether the monopole is on the left or on 
the right of the wall.

Thus, the picture is in 
complete  agreement with the conservation of the total electric
charge.

This picture can be readily generalized for $SU(N)$. Then there 
are
 $N-1$ different monopoles corresponding to $N-1$ (= rank for 
$SU(N)$)
  Abelian $U(1)$ factors in the Cartan subalgebra of $SU(N)$. 
 Repeating the same analysis, one can see that  fundamental and
 antifundamental  representations $[N]$ and $[\bar{N}]$
  are induced   on the  domain wall
  $(n, n+1)$ and $(n+1,n)$ respectively. Here $n = 0,1,... N-1$ 
corresponds to
 $a  = 2\pi n$. One can see that taking two domain walls $(n,
n+1)$  and $(n+1, n+2)$ one can get all representations
 corresponding to the product of two fundamental representations
 $[N] \times [N]$. For $m$ consecutive walls we  have $[N] \times 
[N]... \times [N]$. If one  takes $m=N-1$
 than one can get  antifundamental representation, 
 in full agreement with the fact 
  that the  $N-1$ walls  $(0,1),~(1,2),~... (N-2,N-1)$  make 
together a wall
 $(0, N-1)$,  which is equivalent to the  wall $(N, N-1)$. But the last
 one,  as any $(n+1,n)$ wall,  corresponds to  antifundamental
representation $[\bar{N}]$.

Returning to our original problem, SUSY gluodynamics,
we note that the phase of $\lambda^2$ in a sense plays a role analogous to 
the axion field. It could be interesting to pursue the analogy between
the Abelian toy model and SUSY gluodynamics further.

In summary, in supersymmetric theories which have degenerate 
vacua with very
different physical properties, 
the fact that the confining string can end on a domain wall
is quite natural. 
Actually,  the wall does not have to be BPS saturated to serve as a 
sink
for the chromoelectric flux carried by the string.

Regardless, it is still 
an interesting question whether all possible  BPS saturated walls are 
dynamically realized in SUSY gluodynamics.
In the next section 
we will attempt to address this question. Although we will not be 
able to give a positive proof, we will show that
the straightforward search in the framework of the
Veneziano--Yankielowicz Lagrangians 
for the walls connecting neighboring chirally
asymmetric vacua is in general a dangerous endeavor. As we shall see
the cusp structure of these Lagrangians makes it impossible to decide
this question without additional nontrivial dynamical information.
We will also present a toy version of the underlying phenomenon. 

\section{Glued Potentials}

\subsection{$N$ counting and paradoxes of the wall-building in  the
Veneziano--Yankielowicz Lagrangians}

First, we briefly remind the relevant formalism.
The effective Lagrangian for SUSY gluodynamics was written down a
long time ago \cite{veneziano} and then amended recently
\cite{kovner2}
to properly
incorporate 
the non-anomalous $Z_N$ symmetry.

We will write down the  Lagrangian realizing the anomalous Ward
identities in terms of 
the chiral superfield 
\beq
S = \frac{3}{32\pi^2N}\,\mbox{Tr}\,W^2\, ,
\label{superfs}
\eeq
 namely,
$$
{\cal L} = \frac{1}{4}\int d^4\theta 
N^2\left( \bar S S \right)^{1/3} 
+ 
$$
\beq
\frac{1}{3} \int d^2\theta NS\left( \ln \frac{S^N}{\sigma^N} +2\pi i 
n\right) + 
\frac{1}{3} \int d^2\bar\theta N \bar S\left(  \ln 
\frac{\bar S^N}{\bar\sigma^N} - 2\pi i n\right) \, 
 ,
\label{VYL}
\eeq
where 
$\sigma$ is a numerical parameter,
$$
\sigma = {\rm e}\Lambda^3 {\rm e}^{i\vartheta /N}\, ,
$$
$\Lambda$ is the  scale parameter, a positive number of dimension 
of mass which we will set equal to unity in the following. Please,
note 
the $N$ factors in Eqs. (\ref{superfs}) and (\ref{VYL}).

An important element in the Lagrangian (\ref{VYL}) is an integer-valued  
Lagrange multiplier  $n$. In calculating the partition function and all 
correlation functions the sum over $n$ is implied. 
The variable $n$ takes only integer values and is not a local
field. It does not depend on the space-time coordinates and, 
therefore,
integration over it 
imposes  a global constraint on the 
topological 
charge. It is easy to see that (after the Euclidean rotation) the 
constraint  takes the form
\beq
\label{topch}
\nu \ =\ \frac{1}{32\pi^2}\int d^4x G_{\mu\nu}^a\tilde G_{\mu\nu}^a 
= Z\, .
 \eeq

While the $F$ term in Eq. (\ref{VYL}) is unambiguously fixed, the 
$D$ term
is not specified by the anomalous Ward identities.
We have chosen it in the simplest possible form,  with the numerical
coefficient which gives the correct large $N$ counting.

The extra term 
added to the Lagrangian is clearly supersymmetric  and is also 
invariant
under all global symmetries of the original theory. The 
single-valuedness of the scalar potential and the $Z_N$ 
invariance which were missing in the original Veneziano- Yankielowicz
effective Lagrangian
are restored \footnote{The 
explicit invariance here 
is $Z_N$ rather than the complete $Z_{2N}$ of the original
SUSY gluodynamics, since we have chosen to write our effective 
Lagrangian for 
the superfield which is invariant under $\lambda\rightarrow 
-\lambda$.}.
The chiral phase rotation by the angle $ 2\pi k/N$ with 
integer
$k$ just leads to the shift of $n$ by $k$ units. Since $n$ is summed 
over in the functional integral,
the resulting Lagrangian for $S$ is indeed $Z_N$ 
invariant.
  
The constraint on $[S-\bar S]_F$ following from the Lagrangian 
(\ref{VYL}) results in a peculiar form of the 
scalar potential. 
The expression for the scalar potential
is given in Ref. \cite{kovner1}.

Eliminating, as usual the $F$ component of $S$ with the help of 
classical 
equations of motion at fixed $n$, the effective potential can be 
written as
\beq
U(\phi)\ =\ -V^{-1}\ln\left[
\sum_n\exp\{-4N^2 V (\phi^*\phi)^{2/3}[\ln^2|\phi|+(\alpha +\pi 
n)^2]\}\right]\, .
\label{scalpot}
\eeq
Here $V$ is the total space-time volume of the system, $\phi$
is the lowest component of the superfield $S$, and 
$\alpha={\rm Arg}
 (\phi)$.
In the limit $V\rightarrow\infty$ only one term in the sum over $n$ 
contributes
for every value of $\alpha$; which particular term depends on the 
value of $\alpha$. Thus, for $-\pi/N<\alpha<\pi/N$ the 
only 
contribution comes from $n=0$.
In this sector the scalar potential is
\beq
U(\phi )= 4N^2(\phi^{*}\phi)^{2/3}
\ln \phi \ln \phi^{*}          
\, ,
\eeq
In general we have
\beq
U(\phi ) &=& 4N^2(\phi^*\phi)^{2/3}
\ln (\phi e^{-i{2\pi n/ N}}) \ln (\phi^*e^{i{2\pi n/ N}})
\label{Uphi}
\eeq
$$
\mbox{at}  \ \ \ \ {(2n-1)\pi\over N}< {\rm  arg}\phi
<{(2n+1)\pi\over N}\, .
$$

In other words, the complex $\phi$ plane is divided into $N$ 
sectors. The scalar potential in the $n$-th sector is just that in the 
first sector rotated by $-2\pi/N$. The scalar potential itself is 
continuous,
but its first derivative in the angular direction experiences a jump
at arg$\phi = (2n+1)\pi/N$. The scalar potential is ``glued" out of $N$ 
pieces. 
The $Z_N$ symmetry is explicit in this expression.
It is quite obvious that the problem at hand has $N+1$
supersymmetric minima -- $N$ minima  at $\phi = e^{i{2\pi n/ N}}$, 
corresponding 
to a non-vanishing value of the gluino condensate (spontaneously 
broken discrete chiral symmetry), and a minimum at $\phi = 0$
(unbroken chiral symmetry). 

Including the kinetic term of the field S, as it appears in
Eq. (\ref{VYL}), leads to the following   effective Lagrangian:
\beq
L=N^2\{\partial_\mu \phi^{1/3}\partial_\mu\phi^{*1/3}+U(\phi)\}\, .
\label{effl}
\eeq

We now ask ourselves whether this Lagrangian can be used to find 
an explicit
wall solutions. In fact, the solution for the type I wall  has been
considered in detail in \cite{kovner1} and was found to exist and to
be BPS-saturated. The situation with the type II walls   is more 
complicated.
Note that any field configuration that interpolates between the two
vacua at $\phi=1$ and $\phi=e^{i{2\pi / N}}$ has to go through a
point where the phase of the field $\phi$ is ${\pi / N}$. At this
point the scalar potential has a cusp, and one has to be very careful
in treating such configurations. 

As an illustration,  let us forget for a while
 about possible complications and estimate the tension
of the type II wall  using the Lagrangian (\ref{effl}). The
potential and kinetic energies should contribute to the tension of the
wall roughly the same amount, so we concentrate on the kinetic
term. The variation of the field $\phi$ inside the wall at large $N$ is
$\Delta\phi\sim O(1/N)$. The mass $m$ of the field $\phi$ is 
independent of
$N$ with our choice of the coefficient of the kinetic
term\footnote{This is, in fact,  how the meson masses   should 
behave 
at
large $N$ and this is the reason of choosing the coefficient $N^2$ in
front of the kinetic term in Eq. (\ref{VYL}).}. The width of the wall
obviously is of order $L\sim 1/m \sim O(1) $. The kinetic energy 
contribution is,
therefore,
\beq
E_{\rm kin}\sim N^2 (\Delta\phi)^2/ L^2\sim O(1)\, , \,\,\, 
\varepsilon\sim E_{\rm kin} L \sim O(1)\, .
\label{otsenka}
\eeq
The same estimate is obtained if we consider the potential energy 
contribution.

Surprisingly,  this is  by far lower than the BPS bound on the wall 
energy
in the original theory,  Eq. (\ref{vacen1}), which gives 
$\varepsilon\sim O(N)$.
At  first sight this might seem to be  an arithmetic  paradox even in 
the
framework of the effective theory {\em per se}, viewed as a 
generalized  Wess--Zumino model. 
In the generalized  Wess--Zumino  models with a superpotential 
${\cal 
W}(S)$ 
the BPS
bound is (\cite{dvali1,chibisov}, see also \cite{add})
\beq
\varepsilon > 2|{\cal W}(S_1)-{\cal W}(S_0)|\, ,
\label{bound}
\eeq
where $S_{0,1}$ are the values of the field in two vacua between
which the given wall interpolates. 
Taken at its face value that would give a bound of $\varepsilon\sim
O(N)$ for the type II wall we are considering.

In fact, there is no arithmetic paradox here. The BPS bound (\ref{bound})
on the wall
energy in the generalized  Wess--Zumino  models
assumes that the superpotential is smooth. In the 
effective theory (\ref{VYL}), for the walls that 
cross
the cusp (type II), it experiences a jump,  
$$
\Delta{\cal W}_{\wedge} = {\cal 
W}(S_{\wedge +}) - {\cal 
W}(S_{\wedge -})\neq 0\, .
$$
(Here  $S_\wedge$ is the point where the wall crosses the 
discontinuity
line, and ${\cal W}(S_{\wedge -})$ and ${\cal W}(S_{\wedge+})$ are 
the values of 
the
superpotential below and above this line, respectively.) The 
superpotential which is obtained
from Eq. (\ref{VYL}), after summation over $n$, has a phase 
discontinuity
along
the same lines where the effective scalar potential (\ref{Uphi})
has a cusp.   Accounting for the jump, modifies the bound, and, 
instead 
 of Eq. (\ref{bound}) we, therefore, have 
\beq
2\left| {\cal W}(S_1)- {\cal W}(S_0) - \Delta{\cal W}_{\wedge} \right| 
\, .
\label{bound1}
\eeq
Due to the discontinuity  $\Delta{\cal W}_{\wedge} $, the expression 
in Eq. 
(\ref{bound1}) is $O(1)$, in full accord with Eq. (\ref{otsenka}). 

Nevertheless, even though there is no paradox at  the level of
the effective theory, clearly the effective potential grossly 
misrepresents the tension of the type II wall.
This is  surprising since the field $\phi$ changes slowly
inside the wall, and normally one would think that the effective
potential should properly describe  slowly varying configurations.
It is clear that this failure is intimately connected with the cusp
structure of the effective potential. Our aim now is
 to understand what is the physical origin
of the cusp structure. To warm up we consider first a 
very simple (non-supersymmetric) model, leading to a similar 
structure, and then move on to a more general picture of the 
phenomenon in supersymmetry.

\subsection{Glued 
effective potential in a simple model}

To understand how an effective potential with cusps can appear 
from a smooth potential of the original theory
it is
best to consider an explicit example.
Let us take a (non-supersymmetric)  theory of two scalar fields
with the potential (Fig. 1)
\begin{equation}
U(\phi,\chi)=\frac{\lambda}{2} \, (\phi^2-\eta^2)^2+\frac{\zeta}{2}\, 
(\chi^2-\mu^2)^2-
\kappa^2\phi\chi
\label{origth}
\end{equation}
where $\phi$ and $\chi$ are real fields and $\lambda$ and $\zeta$ 
are the coupling constants. The coupling constant $\kappa$ is taken 
to be real. The mass of the $\phi $ quantum 
$M^2=  4 \lambda\eta^2$, while that of the $\chi$ quantum
$m^2= 4\zeta\mu^2$. 
Let us assume that the field $\phi$ is much heavier than the field 
$\chi$,
 $M
\gg m$. For technical simplicity we will also assume that
$\kappa \ll m,M$, and the expectation values  $\eta ,\mu$ are of 
the same order of magnitude, although this is not crucial. 

\begin{figure}
  \epsfxsize=9cm
  \centerline{\epsfbox{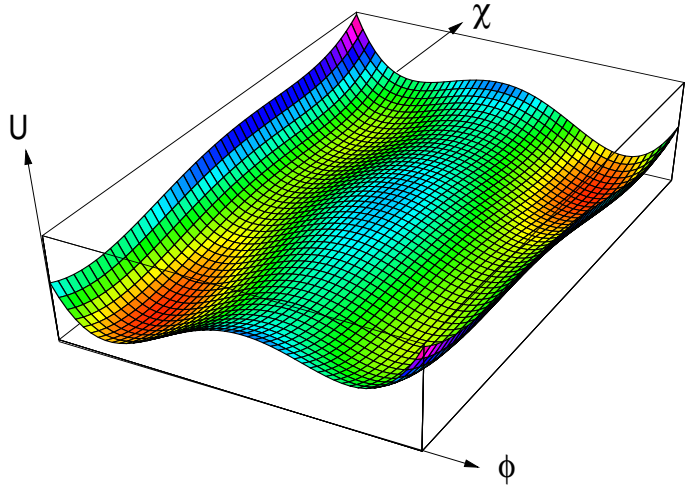}}
  \caption{ The scalar potential in the model considered in Sect. 3.2.}    
\end{figure}

The theory has {\em two} symmetry breaking minima. When both
coupling constants $\lambda ,\zeta $ are large, those are
\beq
\{ \phi= \eta, \ \ \ \chi=\mu \}\,\,\, \mbox{and}\,\,\, 
\{ \phi=-\eta, \ \ \ \chi=-\mu\}\, . 
\label{twovacua}
\eeq
Let us now derive an  effective potential for the field $\chi$ by 
integrating out $\phi$. 
To calculate the effective potential
in the leading adiabatic approximation
we fix the value of $\chi$  and solve for $\phi$ in this background.
Note that for a fixed and not too large 
value of $\chi$ the potential for $\phi$ has two local minima.
Generically, the two local minima are nondegenerate.
For $\chi <0$ the state $\phi=-\eta$ has lower energy \footnote{We 
neglect here
small corrections of order $\mu/\eta$ and $\kappa /\eta$
to the values of $\phi$ at the minima.}, while for
$\chi >0$ the global minimum is at $\phi =\eta$.
At $\chi =0$ both local minima  become degenerate. At this point 
there is 
a
discontinuous change  of  the vacuum in the heavy sector. As a result
the effective potential develops a cusp of precisely the same nature
as discussed in the previous section, 
\begin{equation}
U_{\rm eff}=-\kappa^2\eta|\chi|+\frac{\zeta}{2}(\chi^2-\mu^2)^2\, .
\label{glued}
\end{equation}
Thus, although the underlying theory is perfectly smooth, the 
effective theory
has two sectors and a cusp at $\chi =0$ (Fig. 2).

\begin{figure}
  \epsfxsize=7cm
  \centerline{\epsfbox{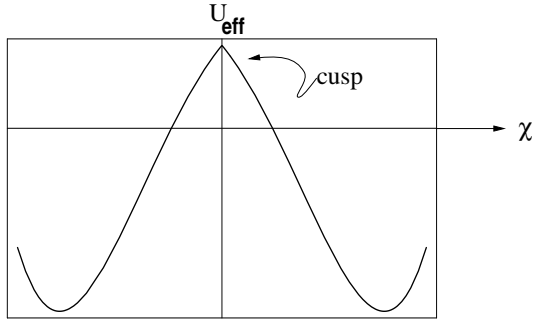}}
  \caption{ The effective  potential obtained after the heavy field 
$\phi$ is integrated out. The cusp at $\chi =0$ reflects a 
restructuring of the vacua in the $\phi$ sector. }    
\end{figure}

We can again pose the question about the existence of the wall, and
the calculability of its energy density from the effective Lagrangian.
Clearly there exists a solution of the original equations of motion,
stemming from Eq. (\ref{origth}), 
with
the wall boundary conditions, i.e. interpolating between the two 
vacua of Eq. (\ref{twovacua}). 
 It is equally clear that the
estimate of its energy from the effective Lagrangian will
be generically incorrect.
The obvious reason is that the effective Lagrangian is completely
independent
of the larger mass scale $M$. This is natural since it was
calculated in the leading adiabatic approximation, i.e. in the limit
$M\rightarrow \infty$.
On the other hand,  to produce the wall one has to excite the heavy
field $\phi$, which jumps from $-\eta$ to $+\eta$ inside the wall 
profile. This costs  energy proportional to $M$, 
so the wall energy density
in this theory must be proportional to $M$. 
The wall tension in the present example can be calculated directly 
from
the ``fundamental'' Lagrangian (\ref{origth}) without appealing to the 
effective
Lagrangian (\ref{glued}).
Roughly it behaves as
\beq
\varepsilon=xM\eta^2+ym\mu^2
\eeq
where $x$ and $y$ are numbers of order one.
If the expectation values of the heavy and light fields are of the 
same
order, $\eta\sim\mu$, the bulk of the wall energy is contributed by 
the
heavy modes. In that case the wall energy can not be
obtained from the effective potential for $\chi$. One can say that
a dominant part of the wall tension is associated with the cusp.

Physically the picture of what is happening
is very simple. The field $\chi$ is light and
therefore changes slowly inside the wall, on the scale $1/m$.
The heavy field $\phi$ follows this change adiabatically almost
everywhere in space, except for the region where $\chi=0$.
In this region, within a distance of order $1/M$, the value of $\phi$ 
changes
from $-\eta$ to $\eta$. The big contribution to the wall energy
density, proportional to $M$ comes precisely from this small region 
in
space in which the field $\chi$ sits on the cusp of the effective 
potential.

If we use the effective potential to calculate the wall energy, the
result will be of order $m\mu^2$, since this is indeed the
contribution of the light field. The 
contribution of the heavy fields can be thought of as an extra
contribution of the cusp in the light field effective potential.

There are several lessons we want to draw from this toy model.
First, the physical reason for the appearance of the cusp in the 
effective potential (glued potential) at some value $\chi_0$ of the
field is that at this particular value
of the light field, the system of heavy fields has two (or more) 
degenerate
ground states. In general, when the value of the light field changes
continuously, the heavy fields follow adiabatically after this change.
However when the light field passes through  $\chi_0$, there is a 
level
crossing in the heavy system and the properties of the heavy field 
vacuum
change discontinuously. This ``first order phase transition"
leads to discontinuity in the first derivative of the effective potential,
and, therefore, a cusp.

The second lesson, is that the wall configuration that connects the 
points
on different sides of the cusp necessarily involves excitation of heavy
modes. This is so since close to the point $\chi_0$ inside the wall
the heavy modes {\it must} rearrange in order to make the transition 
between
the two degenerate vacuum states. 

Finally, the description of the wall with the help of naive glued 
potential
is not valid if the bulk of the wall energy comes from the change of
the heavy fields. 
The problem stems from the fact that the effective potential is 
calculated
in expansion
in powers of $1/M$, which is the adiabatic approximation. As we 
already 
stressed, the adiabatic approximation breaks down inside the wall 
due
to the level crossing. On one hand, this is a signal 
of possible appearance of
terms proportional $M$ in the effective action.
On the other hand, this is precisely the situation 
in which a nontrivial topological
Berry phase should appear.
A more careful calculation of
effective action should reveal the presence of the terms of the type
\beq
 M[\phi(\infty)-\phi(-\infty)]^2\, .
\eeq
This topological term does not affect the vacuum sector, but adds the
missing large piece, cusp contribution to the energy of the wall.

Returning to SUSY gluodynamics it is now clear why the original BPS
bound is so badly violated by a wall configurations in the effective
theory.
The reason is that the effective theory misses a large cusp
contribution to the energy which comes from the heavy modes not
appearing in the effective Lagrangian that are excited inside the
wall. We conclude, therefore,  that the Veneziano--Yankielowicz 
effective Lagrangian, as it
stands,
can not be used to calculate the wall energy; without
additional
information we can not say whether or not the BPS saturated type II 
walls  exist in SUSY gluodynamics. 

In the next section we would like to give a detailed example of how 
this
situation arises in a supersymmetric theory. The model we will
consider has the same symmetries as SUSY gluodynamics and an 
effective
potential of the Veneziano--Yankielowicz type.
We will see in detail how the cusp structure of the effective
Lagrangian appears when integrating out the heavy superfields and 
will
be able to trace exactly the missing piece of the wall energy.

\subsection{Supersymmetric model with $N$ sector superpotential}

The model of the previous subsection was only intended for 
explaining how cusps arise in the effective potential.
We now want to consider 
a model which captures more features inherent to SUSY 
gluodynamics. Consider a
generalized Wess-Zumino model of two scalar chiral superfields 
$\Phi$ and $X$, 
with the superpotential
\beq
{\cal W} ={a\over N+2}\Phi^{N+2}- b\Phi X+{c\over 2} X^2\, ,
\label{superpo}
\eeq
where the coefficients $a,b$ and $c$ are real  positive numbers.
This model  obviously has a $Z_N$ symmetry, under which
\beq
\Phi\rightarrow \Phi e^{i{2\pi n / N}}\, ,\, \, \, 
X \rightarrow  X e^{i{2\pi n /  N}}\, ,\, \, \,  \theta\rightarrow
\theta e^{i{2\pi n /  N}}\, . 
\eeq
The global minima of the energy are determined from the equations
\beq
{\partial {\cal W}\over\partial\Phi}=a\Phi^{N+1}-bX=0\, , \,\,\,  
{\partial {\cal W}\over\partial X}=-b\Phi+c X =0 \, .
\label{minima}
\eeq
These equations have one $Z_N$ symmetric solution
\beq
\Phi=X =0
\eeq
and $N$ solutions which spontaneously break the $Z_N$ symmetry
\beq
\Phi_n=\left({b^2\over ac}\right)^{1/N}e^{i{2\pi n / N}},\ \ \ \ \
X _n={b\over c}\, \Phi_n\, .
\label{reshen}
\eeq
Choosing $b\gg c$, we find that the field $\Phi$ near the asymmetric 
vacua
is very heavy; its  mass 
\beq
M=(N+1)\, {b^2\over c}\, ,
\label{mahi}
\eeq
 while 
the field $X$ is light,  with the mass $m=c$ \footnote{Strictly 
speaking, the mass matrix is non-diagonal. There is a small 
admixture of $X$ in the heavy diagonal combination, and a small 
admixture of $\Phi$ in the light diagonal combination. These 
admixtures are $O(cb^{-1}N^{-1})$, and the absolute shift in the mass 
eigenvalues is $O(cN^{-1})$. This shift, as well as the mixing 
mentioned above, can be neglected in the limit
$N\gg 1$ and $b\gg c$. Moreover, if $N\gg 1$, 
it is  not necessary to require $b\gg c$. Even at $b\sim c$ the  
hierarchy of masses takes place, $M\gg m$, and the off-diagonal 
elements of the mass matrix are negligible.}.
 
The  effective potential for the field $X$ is obtained
by eliminating the heavy field $\Phi$ by virtue $\partial {\cal W}/ 
\partial\Phi =0$, see 
 the first equation in
(\ref{minima}).
The condition $\partial {\cal W}/ 
\partial\Phi =0$  has $N+1$ solutions
\beq
\Phi_*(X)=\left({b\over a}X\right)^{1\over N+1}\exp\left\{ {2\pi  i 
n\over 
N+1}\right\}\, .
\label{solutions}
\eeq
The solution has to be substituted in the superpotential,
$$
{\cal W}[\Phi, X] \rightarrow {\cal W}[\Phi_*(X), X]
=
$$
\beq
-\frac{N+1}{N+2}b\left( \frac{b}{a}\right)^{1/(N+1)} X^{(N+2)/(N+1)}
\exp\left( \frac{2\pi i n}{N+1}\right) +\frac{c}{2}X^2\, .
\label{efsuperp}
\eeq
This is not the end of the story, however. 
The effective superpotential is obtained by choosing for every value 
of
$X$ the solution that gives a minimal energy.
The energy as a function of the lowest component $\chi$ of the
superfield $X$ for each   branch  is
\beq
\left| -b\left({b\over a}\chi\right)^{1\over N+1}\exp\left\{ {2\pi i
n\over
N+1}\right\} +c\chi\right|^2\, .
\label{tben}
\eeq
Clearly, the energy is minimal for the branch for which  Arg$\chi$ is
closest to 
$$
{2\pi n\over N+1}+{1\over N+1}{\rm Arg}\,\chi\, .
$$
For instance, if $\chi$ is real and positive, $n=0$,
if Arg$\chi = 2\pi / N$ then $n=1$, and so on. At Arg$\chi = \pi  / 
N$ both branches, with $n=0$ and $n=1$, have the same energies.
The resulting effective superpotential has $N$ sectors and is ``glued" 
along $N$ rays, 
\beq
{\cal W}_{\rm eff}={c\over 2}X^2-{N+1\over N+2}\, b\left({b\over
a}\right)^{1\over N+1} X^{{(N+2)}/{(N+1)}}\, ,
\label{toypot}
\eeq
$$
\mbox{at} \,\,\,  -{\pi\over N}<{\rm Arg}\chi < {\pi\over N}\, ,
$$
\beq
{\cal W}_{\rm eff}={c\over 2}X^2-{N+1\over N+2}\, b\left({b\over
a}\right)^{1\over N+1}\exp\left\{ {2\pi i\over
N+1}\right\} X^{{(N+2)}/{(N+1)}}\, ,
\label{toypot1}
\eeq
$$
\mbox{at} \,\,\,  {\pi\over N}<{\rm Arg}\chi <{3\pi\over N}\, ,
$$
and so on (see Eq. (\ref{efsuperp})). 
The discontinuities in the effective superpotential (or, equivalently, 
the
cusps in the effective potential) occur along the rays
$$
{\rm Arg}\chi=\frac{(2k+1)\pi}{ N}
$$
 where  two branches
in Eq. (\ref{solutions}) with $n=k$ and $n=k+1$ are degenerate in 
energy, see Eq. (\ref{tben}).
For instance, at Arg$\chi = \pi /N$
$$
{\cal W}_{\rm eff} (\chi_{\wedge\pm})=
e^{2\pi i /N}\left[
\frac{c|\chi |^2}{2} -\frac{N+1}{N+2} b\, \left( 
\frac{b}{a}\right)^{\frac{1}{N+1}}|\chi |^{(N+2)/(N+1)}\exp\left( \pm 
\frac{\pi i}{N+1}\right)
\right]\, .
$$

Now we are ready to address the issue of the domain walls. 
Consider a domain wall that connects two adjacent
asymmetric vacua. The BPS bound on its tension  is
\beq
\varepsilon>2|{\cal W}(\Phi_0,X_0)-{\cal W}(\Phi_1,X_1)|={2N\over
N+2}{b^2\over c}
\left({b^2\over ac}\right)^{2\over N}\, \sin \left( {2\pi\over 
N}\right)\, .
\label{granitsa}
\eeq
At large $N$ this is of order $1/N$. 
On the other hand, a naive estimate based on the effective potential 
for the light  field  would yield
\beq
\varepsilon\sim m\left| \Delta \chi \right|^2\sim 
{b^2\over c}
\left({b^2\over ac}\right)^{2\over N}\, \frac{1}{N^2}\sim N^{-2}
\, .
\label{epots}
\eeq

Just like in SUSY gluodynamics the two expressions are
incompatible. We know already that the reason is that the adiabatic
approximation used to derive the effective Lagrangian breaks down 
at
the cusp. For configurations  crossing the cusp an extra
``topological'' term has to
be added to the effective Lagrangian, as discussed in the previous
section. 
Let us first estimate a part of the tension associated with the cusp,
and the corresponding restructuring of the heavy field
$\phi$. A straightforward estimate analogous to that of Eq. 
(\ref{epots}) is
\beq
\varepsilon_\wedge \sim M|\Delta \phi |^2 \sim {b^2\over c}
\left({b^2\over ac}\right)^{2\over N}\, \frac{1}{N}
\, ,
\label{epcuspots}
\eeq
i.e.  we reproduce the order of magnitude of the BPS bound
(\ref{granitsa}). 

As a matter of fact, in the  present case one can do better than that. 
To  this end we note, that the only point where the adiabatic
approximation breaks down is at the cusp. In other words, almost
everywhere throughout the space the heavy field $\phi$ does indeed
follow the change of $\chi$ adiabatically. Only at the point inside
the wall, where 
$$
\chi=\rho e^{i\pi /  N}\,  ,
$$
the value of $\phi$
changes
rapidly. This change, of course, does not happen abruptly, but rather 
on
the scale of the inverse mass $M^{-1}$ of the field $\phi$. The
field $\chi$ remains constant throughout the region of space where 
the
rapid variation of $\phi$ takes place. The profile of $\phi$ in this
region as well as the energy associated with this variation can be
calculated  by considering the original theory at a frozen cusp value 
of $\chi$.

  If $\chi=\rho \exp{(i\pi /  N)}$, the wall profile of the field $\phi$
is  determined from  the following
superpotential
\beq
{\cal W}_\phi={a\over N+2}\Phi^{N+2}-b\rho e^{i\pi / N}
\Phi\, .
\label{wphi}
\eeq
At $\chi=\rho \exp{(i\pi /  N)}$ the two branches of
Eq. (\ref{solutions}) are degenerate.
The $\phi$ wall under consideration  interpolates between
\beq
\phi_{*0,1}=\left( {b\over a}\rho\right)^{1\over N+1}\, \exp
\left[ i\pi \left( \frac{1}{N} \pm\frac{1}{N+1} \right)\right]\, ,
\eeq
where the upper and lower signs correspond to the 
final and initial points, respectively.
These points are  two degenerate minima of the potential 
stemming from Eq. (\ref{wphi}).
The $\phi$ wall in question is BPS saturated. This is
because the BPS equation in this case is a pair of the first order
equations for two real fields (real and imaginary parts of $\phi$),
\beq
\partial_z \phi =\frac{d\bar{{\cal W}}}{ d\bar\phi}\, \exp\left( 
-\frac{i\pi}{2}+ \frac{2\pi i}{N}\right)\, ,
\label{bpseq}
\eeq
which possess one conserved quantity (see Ref. \cite{chibisov} for 
details)
\beq
{\rm Im}\, \left[ ({\cal W}(\Phi)-{\cal W}(\Phi_{*0}))\exp\left( 
\frac{i\pi}{2}- \frac{2\pi i}{N}\right)\right]\, =0\, .
\label{conserv}
\eeq
Using  Eq. (\ref{conserv}) one  can always eliminate one real field,
getting in this way a trajectory in the plane $\{ \mbox{Re}\phi\,  , \, 
\mbox{Im }\phi\}$ that connects $\phi_{*0,1}$. The trajectory 
depends on one real variable. The
resulting one first-order equation for one real function always has a
solution.

We conclude that the tension of the $\phi$ wall (the cusp term) is 
given by the BPS bound for the theory with the superpotential 
 (\ref{wphi}),
\beq
\varepsilon_\wedge =4\, {N+1\over N+2}\, (b\rho)^{(N+2)/(N+1)}a^{-
{1/
(N+1)}}
\sin{\pi\over N+1}\, .
\label{cuspten}
\eeq

This is precisely the term  that has to be added to the effective
potential of $\chi $ in order to be able to calculate the wall tension
properly.
Equation (\ref{cuspten}) can be  generalized to a wall configuration 
which
connects any two vacua, and not necessarily the two adjacent ones. 
The
extra term in this case would be just the sum of contributions of all  
cusps crossed by 
 the wall.

Note that although the $\phi$ component is built above through the 
BPS saturation, the full $\{ \phi \, , \, \chi\}$ wall in the original 
theory is not BPS-saturated, at least, in some range of parameters.
(We mean the type II wall 
 connecting, say, two neighboring asymmetric vacua with $n=0$ and 
$n=1$ in Eq. (\ref{reshen}).)

To see that this is indeed the case consider the model (\ref{superpo})
with $a=b=c=1$ and $N\gg 1$. (For these values of parameters
the field $\Phi$ is still much heavier than $X$, see the footnote 
following Eq. (\ref{mahi})).  Assume that the wall is BPS saturated.
Then
\beq
\partial_z \phi = \left( \bar{\phi}^{N+1} -\bar\chi \right)\, \exp\left( 
\frac{i\pi}{2} + \frac{2i \pi}{N}\right)\, , \,\,\, 
\partial_z \chi = \left(- \bar\phi + \bar\chi \right)\, \exp\left( 
\frac{i\pi}{2} + \frac{2i \pi}{N}\right)\, . 
\eeq
As a consequence,
\beq
\partial_z \phi + \partial_z \chi = \left( \bar{\phi}^{N+1} -\bar\phi 
\right)\, \exp\left( 
\frac{i\pi}{2} + \frac{2i \pi}{N}\right)\, ,
\eeq
where the right-hand side contains no dependence on $\bar\chi$.
We can take advantage of this fact. Multiply both sides
by $\partial_z \bar\phi$ and integrate over $z$ from $-\infty$ to 
$+\infty$. Then
\beq 
\int_{-\infty}^{\infty} (\partial_z \phi + \partial_z \chi )\partial_z 
\bar\phi dz =\exp\left( 
\frac{i\pi}{2} + \frac{2i \pi}{N}\right) \,\left[ 
\frac{1}{N+2}{\bar\phi}^{N+2} - 
\frac{1}{2} 
{\bar\phi}^{2}\right]_{-\infty}^{\infty}   =
-\frac{\varepsilon_{\rm BPS}}{2}\, ,
\label{promez}
\eeq
where $\varepsilon_{\rm BPS}$ is the tension defined on the 
right-hand side of Eq. (\ref{granitsa})
and the values of the $\phi$ field in the vacua, Eq. (\ref{reshen}), are 
substituted. 
Equation (\ref{promez}) implies
\beq
\int_{-\infty}^{\infty} ( \partial_z \chi)(\partial_z \bar\phi )dz 
= -\frac{\varepsilon_{\rm BPS}}{2} - C_1
\label{promez1}
\eeq
where $C_1$ is a real {\em positive} number.
On the other hand, 
$$
C_2 = \int_{-\infty}^{\infty} (\partial_z \phi + \partial_z \chi 
)(\partial_z \bar\phi + \partial_z \bar\chi ) dz  = \int_{-
\infty}^{\infty} (\partial_z \phi \partial_z \bar\phi + \partial_z \chi 
\partial_z \bar\chi )
+
$$
\beq
 \left[  \int_{-\infty}^{\infty} ( \partial_z \chi)(\partial_z \bar\phi 
)dz  +\mbox{H.c.}\right] 
\eeq
where $C_2$ is a real {\em positive} number. The first term is 
obviously 
equal to $\varepsilon_{\rm BPS} / 2$, while the second term can be 
read off from Eq. (\ref{promez1}). In this way we obtain
\beq
C_2 = \frac{\varepsilon_{\rm BPS}}{2} + 2\left(  
-\frac{\varepsilon_{\rm BPS}}{2} - C_1\right)\, ,
\eeq
or $C_2 + 2C_1   = -\varepsilon_{\rm BPS}/2$. 
This relation is obviously inconsistent, which proves that
we cannot built a BPS saturated wall in the problem at hand.

Thus,  the lesson to be drawn is as follows. 
Consideration of the effective low-energy theory, by itself,
yields information on the vacuum structure. The fact of existence of 
the walls can be unambiguously inferred from this information.
But neither the nature of the wall (BPS {\em versus} non-BPS), nor
its tension can be properly found from the analysis
 of the effective low-energy theory if the corresponding potential is 
glued from distinct sectors, and the wall in question crosses the 
cusps. 

\subsection{Elements of the general theory}

Given an effective low-energy theory, obtained after integrating out 
all heavy fields,  with a discrete set of
degenerate vacua (as it is typical for supersymmetry),
the question we ask is:   can one infer from this 
low-energy theory the existence of the BPS walls interpolating 
between the distinct vacua? Under what circumstances
 the BPS wall seen in the effective theory is a reflection of the wall in 
the full theory?
And {\em vice versa}, if we see no BPS walls in the effective 
low-energy 
theory
does it mean there are no such walls in the full theory?

The full general theory is not yet worked out, and the answers to 
these 
questions in the generic situation are not known so far.
In this section we will present some illustrative considerations
which are valid in the simplest possible setting: the generalized   
Wess-Zumino models, with all  parameters in the superpotentials
that are real. We will limit ourselves to the  wall solutions
where all fields take real values, so we do not have to travel in 
the complex plane, and can apply a rich physical intuition stemming 
from the fact that the BPS equations in this case are those of 
high-viscosity fluid (the so called creek equations) \cite{chibisov}.
 We will see that even in this simplest case the situation 
is quite non-trivial. Whenever the low-energy theory
has a glued potential, we can  count the number of 
distinct walls but,  generically, can say nothing about their BPS 
nature and/or tension. 

Let us consider  for simplicity two chiral superfields, $\Phi$ and $X$,
and a superpotential shown on Fig. 3. (More exactly, Fig. 3
displays ${\cal W}$ as a function of $\phi,\chi$
for real values of $\phi,\chi$.)
Shown are two ``mountain ridges",
the left ridge and the right one, separated by a ``canyon".
The heavy field is $\Phi$, the light one is  $X$. The vacua of the 
theory correspond to the points
where
\beq
\frac{\partial{\cal W}}{\partial\Phi} =0\, , \,\,\, 
\frac{\partial{\cal W}}{\partial X} =0\, .
\label{AA}
\eeq
These points are denoted by $A,B,C,D,E$ on Fig. 4. The points $A,B,C$
lie on the left ridge, the point $D$ on the right ridge while the
point $E$ belongs to the bottom of the canyon.

\begin{figure}
  \epsfxsize=14cm
  \centerline{\epsfbox{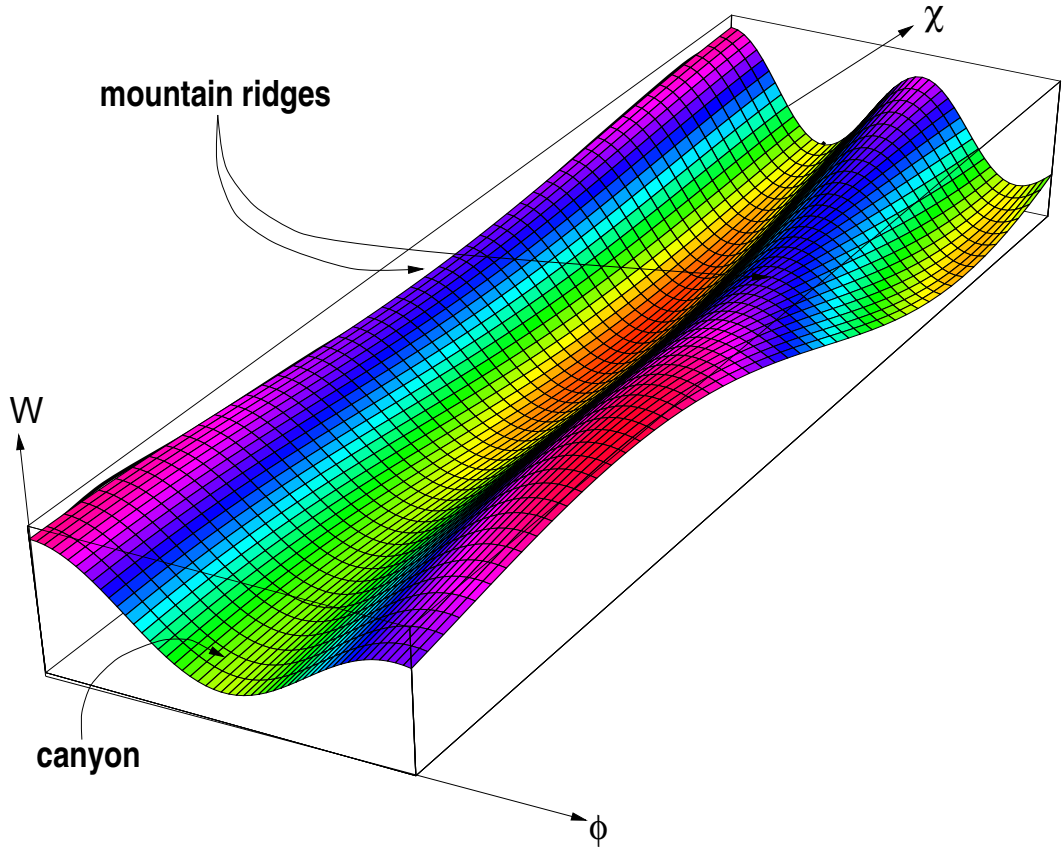}}
  \caption{ A superpotential with two mountain ridges and a canyon
and five vacuum states.}    
\end{figure}

\begin{figure}
  \epsfxsize=13cm
  \centerline{\epsfbox{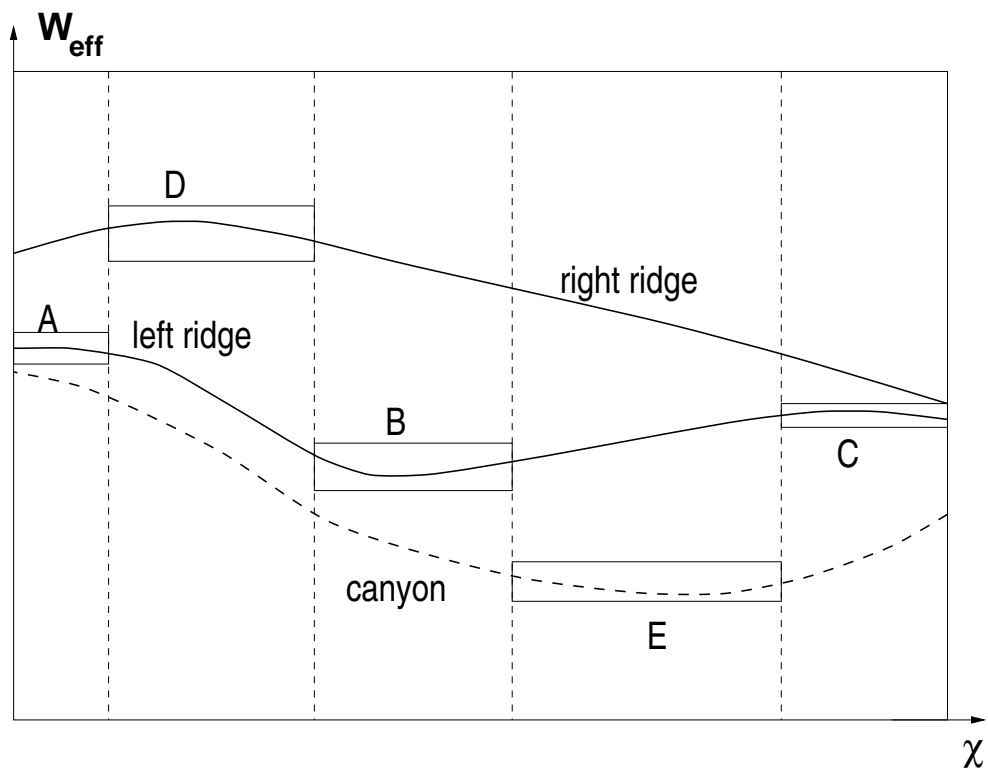}}
  \caption{ The projection of the superpotential of Fig. 3 onto the
${\cal W}\chi$ plane. Shown are the shapes of the mountain ridges
and the canyon bottom. The points of extrema in ${\cal W}$ are 
denoted by $A,B,C,D,E$. }    
\end{figure}

The low-energy reduction is obtained by eliminating the field
$\Phi$
by virtue of the equation
\beq
\frac{\partial{\cal W}(\Phi , X)}{\partial\Phi} =0
\,\,\, \rightarrow \,\,\,  \Phi = \Phi_* (X)\, .
\label{A}
\eeq
Substituting $\Phi_* (X) $ back in the superpotential ${\cal W}$
we get an effective low-energy superpotential
\beq
{\cal W}_{\rm eff}( X) = {\cal W}(\Phi_* (X) , X)\, .
\label{B}
\eeq
Equation (\ref{A}) determines the positions of the mountain ridges
and the bottom of the canyon, while Eq. (\ref{B})
projects them onto the ${\cal W}\chi$ plane. 
One can visualize ${\cal W}_{\rm eff}$ as shadows of the
ridges left by a parallel horizontal beam of light in the 
$\phi$ direction. 

First of all, let us prove that the vacua of the full theory (extrema
of ${\cal W}$) lie on the mountain ridges  and/or bottoms of the 
canyons
(extrema
of ${\cal W}_{\rm eff}$ ).
Indeed,
$$
\frac{d {\cal W}_{\rm eff}}{d X} =
\left[ \frac{\partial{\cal W}(\Phi , X)}{\partial X } 
+ \frac{\partial{\cal W}(\Phi , X)}{\partial\Phi} \, 
\frac{\partial\Phi_*}{\partial X}
\right ] _{\Phi = \Phi_*(X)} = 0\, ,
$$
if Eq. (\ref{AA}) is satisfied. Thus, the extrema of ${\cal W}_{\rm 
eff}$ and ${\cal W}$ coincide.

Let us assume now, for a short while, that the right mountain ridge 
and the canyon do not exist, and the profile of ${\cal W}$ has only 
the left mountain ridge, with three vacua, $A,B$ and $C$.
In this case the solution of Eq. (\ref{A}) is unique.
The full theory has two BPS walls: $AB$ and $CB$. The potential
of the low-energy effective theory ${\cal W}_{\rm eff}( X) $
is a smooth function, and the low-energy theory  has two BPS 
walls too. Their existence can be seen from the creek
equation in the low-energy theory {\em per se},
\beq
\partial_z X = \frac{d {\cal W}_{\rm eff}}{d X}\, .
\label{BB}
\eeq
The wall profile $ X(z)$ we find from Eq. (\ref{BB})
is a little bit different from  what one would get by solving
the creek equations in the full theory,
\beq
\partial_z X =\frac{\partial{\cal W}(\Phi , X)}{\partial X } \, , \,\,\,
\partial_z \Phi =\frac{\partial{\cal W}(\Phi , X)}{\partial\Phi } \, .
\label{creq}
\eeq
The difference vanishes in the limit when the field $\Phi$ becomes 
infinitely heavy; it  dies off as positive powers of $1/M_\phi$.
At the same time , the wall tension found in the effective theory
{\em exactly coincides} with that one would find in the full theory.
No $1/M_\phi$ corrections can appear. 
This is a remarkable  feature of BPS supersymmetric walls.
The tension of such a wall is exactly
determined by the
central charge \cite{dvali1,chibisov,kovner1} which reduces, in turn,
to $|{\cal W}(\Phi_0 , X_0)-{\cal W}(\Phi_1 , X_1)|$
in the full theory, and to $|{\cal W}( X_0)_{\rm eff}-{\cal W}( 
X_1)_{\rm eff}|$ in the effective low-energy theory. 
(Here $\{ \Phi_{0,1}, X_{0,1}\} $ are the points of extrema.) The two 
expressions above coincide {\em identically}. 

Summarizing, if for all values of the light fields
the solution for the heavy fields, to be integrated out, is unique,
the existence of a BPS wall in the effective theory entails
the existence of such a wall in the full theory, and {\em vice versa}.
Moreover,  if one calculates the wall tension in the effective theory, 
one gets the exact answer valid in the full theory, with no
 $1/M_\phi$ corrections.

Let us return  to the superpotential depicted on Fig. 3, with two 
mountain ridges and one canyon. In addition to the 
$AB$ and $CB$  walls, the full theory
has a $BE$ wall which is BPS saturated,
and three continuous sets of BPS walls $AE$, $CE$ and $DE$.
Each set includes an infinite amount of degenerate walls
(by degenerate we mean that the tensions of all walls inside each set 
are the same). The phenomenon of continuously degenerate
supersymmetric walls was first observed in Ref. \cite{SV}.
Besides these BPS walls, the full theory 
may have $AD,BD$ and $CD$ walls that are not BPS saturated.
Depending on the values of parameters in the superpotential it may 
be expedient for some or all non-BPS walls to decay into a pair of 
BPS walls.

What can be said about the effective theory? The corresponding 
low-energy effective potential will be glued out of five pieces, as 
indicated in Fig. 4. In each of  five domains, the branch of the 
effective superpotential corresponding to the  lowest energy
is shown by rectangles. We get a typical five-sector structure
of the scalar potential in the effective theory.
>From this structure we conclude, with certainty, that
the theory under consideration has five degenerate vacuum states, so 
that each pair can be connected by a domain wall. Inspecting the 
low-energy theory, without additional information on the full theory, 
we can say nothing, generally speaking, as to the nature of these 
walls. The reason is obvious: we have no idea where each sector 
comes from,  whether or not two extrema in question belong to one 
and the same mountain ridge (canyon).  If they do, no restructuring
of the heavy field $\phi$ vacuum occurs inside the wall,
and we return back to the situation with the unique solution of Eq. 
(\ref{A}) discussed above.

If extrema from different sectors actually do belong to distinct
mountain ridges or  canyons \footnote{Generally speaking,
distinct sectors can belong to one and the same branch, see points 
$A$ and $C$ in Fig. 4.} (Eq. (\ref{A}) has more than one 
solution) there is no unambiguous way to decide BPS {\em versus} 
non-BPS from inspection of the low-energy theory alone.
We need additional information regarding what happens
with the heavy fields inside the wall. In any case, a part of the wall 
tension associated with the light fields will not saturate 
$\varepsilon$.

\section{The Kaluza-Klein Domain Wall}

As was discussed above, the tension of the walls
interpolating between the neighboring vacua in supersymmetric 
gluodynamics is expected to scale as $N$ rather than $N^2$.
This gives rise to a natural identification of these walls
with the $D$-brane solutions found in Ref. \cite{witten}.
Here we will show, that this phenomenon, ``abnormal" $N$ 
dependence, is actually more general, and shows up
in other wall configurations   related to Witten's analysis.  In fact, the 
low-energy limit of the theory considered
in 
Ref.  \cite{witten} is a  five-dimensional
  Kaluza-Klein (KK) theory with a five-dimensional $SU(N)$ gauge
field $A_{M},~M=1,..,5$ and charged matter.
  After compactification there are two types
 of gauge fields --
 our original $SU(N)$ gauge field $A_{\mu}$ and a new $U(1)$
 gauge field $B_{\mu}$ coming from 
 the  $G_{\mu 5}$ components of the metric tensor as well as a scalar made
 from the fifth component $A_{5}$ of the $SU(N)$ gauge field.
  We are going to demonstrate  that if this theory is modified, so that the
  supersymmetry is broken explicitly,
  there  is a  new type of  domain wall due to the field $A_{5}$.
   By analyzing this low-energy theory {\em 
per se}, with no reference to $D$-branes, we demonstrate that the 
wall
tension scales as $N$, in parallel with the brane-based 
derivation of Ref. \cite{witten}. Moreover, these walls carry 
an  induced fractional charge \cite{kogan}.
Conceptually the situation reminds that with the axion wall discussed
in Sect. 2.2.  

The theory to be considered is gravity plus the gauge field $A_{M}$  
in five dimensions. To warm up we start from the Abelian case,
i.e. the $U(1)$ gauge group. At this stage we also omit the superpartners 
from the discussion.
  The action is
\bq
S = \int d^{5}x \sqrt{-G}\left[ -
\kappa_{5} R - \frac{1}{4e^{2}_{5}}F_{MN}F^{MN} 
+ G^{MN}(\partial_{M}-
iA_{M})\Phi^{\ast}(\partial_{N}+iA_{N})\Phi\right] ,
\label{5action}
\eq
where $\kappa_{5}$ and $e_{5}$ are the 
five-dimensional
gravitational and gauge coupling constants, $G_{MN}$ is the 
metric, $R$ stands for the curvature,  $F^{MN} $ is the gauge field 
strength tensor, all capital Latin 
letters run from 1 to 5, say, $M,N = 1, ...,  5$, 
while the Greek letters 
 $\mu ,\nu \, , ... = 1, ... ,4$.
It is assumed that one of the five dimensions forms a circle, so
that we deal with $M^{4} \times S^{1}$
KK model. The mater sector  consists of 
charged 
scalars $\Phi$,  the simplest possible choice.

After $M^{4} \times S^{1}$ decomposition of the metric 
\bq
  G_{MN} = \left(\begin{array}{lc}
g_{\mu\nu} &  B_{\mu} \\
B_{\nu} & 1 + B_{\mu}B^{\mu} \end{array} \right) ;\;
G^{MN} = \left(\begin{array}{lc}
g^{\mu\nu}+B^{\mu}B^{\nu} &  -B^{\mu} \\
-B^{\nu} & 1  \end{array} \right) 
\eq
we get four-dimensional gravity, two $U(1)$ gauge  fields 
$A_{\mu}$ and 
$B_{\mu}$, plus a  scalar $A_{5}$ (we put dilaton $G^{55}
 = 1$).

For any manifold $K$
 with a  nontrivial $\pi_{1}(K)$ the KK theory contains
special Wilson line operators 
\bq U_{\gamma}
= P \, exp \,\left( i\oint_{\gamma} A dx \right)
\, ,
\eq
where $\gamma$ is 
a closed noncontractible 
contour on $K$. 
In our  case $K = M^{4} \times S^{1}$ and    
 $\pi_{1}(S^{1}) = Z$.  For the $U(1)$ gauge field
$$
U = e^{i\phi}, \;\;\;\; \phi\in [0, 2\pi)\, .
$$
The phase $\phi$ represents the 
constant component of the
 gauge field $A_{5} = [0, 1/R]$, where $R$ is the compactification 
radius.
Other values of  $A_{5}$ can be gauged into the interval $[0, 
1/R)$ 
while  $A_{5} = 1/R$ is gauge equivalent to $A_{5} = 0$. The 
corresponding 
 gauge function is
 $\epsilon = x_{5}/R$.
 This means that the charged (with 
 respect to field $A$)
 fields are related by a gauge transformation,
\bq
\Psi_{A_{5}=1/R}(x,x_{5}) = e^{ix_{5}/R}\Psi_{A_{5}=0}(x,x_{5})
\label{gtr}
\eq
where $x$  denotes four noncompact coordinates.  
Using the standard KK decomposition
\bq
\Psi(x,x_{5}) = \sum_{n=-\infty}^{\infty} e^{inx_{5}/R}\Psi_{n}(x)\, ,
\eq
where $\Psi_{n}(x)$ describes  different four-dimensional fields
with masses $m^{2}_{n} = n^{2}/R^{2}$ and charges 
\footnote{ We
 refer to the charge $q$ as  the KK charge,
as opposed to   the conventional charge defined with  
respect to $A_{\mu}$.}  $q = n$ with 
respect to the KK
gauge field $B_{\mu} = G_{\mu 5}$,  we see that the gauge
transformation (\ref{gtr}) shifts $n
\rightarrow n+1$: \bq
\Psi_{n}(x)|_{A_{5} = 1/R} = \Psi_{n+1}(x)|_{A_{5} = 0}
\eq

Changing adiabatically $A_{5}$ from zero to $1/R$ all levels in the 
particle spectrum are shifted by one.
 For example, a   massless
neutral (with respect to $B_{\mu}$) particle will be transmuted 
into a heavy ($m =
1/R$) charged ($q = 1$) particle. Thus, if  there is a domain
 wall $A_{5}(x)$ interpolating between $A_{5}(-\infty) = 0$ and 
$A_{5}(\infty) = 1/R$, and  one 
 scatters  a  massless neutral particle  with $n=0$ on the 
 domain wall it either reflects or becomes massive and charged.
  The total KK charge must be conserved in the process, since
it is the  gauge charge which generates a part of the general coordinate
 transformation, $\delta G_{MN} = \partial_{M}\epsilon_{N}(x,x_{5}) +
\partial_{N}\epsilon_{M}(x,x_{5})\,$  
which for $\epsilon_{M}(x,x_{5}) = \delta_{M,5}\epsilon(x)$ 
 is the 
$U(1)$
 gauge transformation, $\delta B_{\mu} = \delta G_{\mu 5} = 
\partial_{\mu}\epsilon$. 

The conservation of the KK charge is insured in a way very similar
to the one discussed in Sect. 2.2.
In the presence of the charged   
particles the domain wall itself acquires an  induced KK charge $q$. 
The total charge of the domain wall plus the particle is conserved. 
The real process of the particle penetration through the domain wall
looks as follows: the initial state is the KK neutral 
particle($q=n=0$) 
plus the   charged domain wall, with the charge $q_{i}=+1/2$. The 
final state (if the particle initially 
has momentum $p_{x} > 1/R$) is the charged
particle $q=n=1$ plus the domain wall with the  charge $q_{f} = 
-1/2$. The total charge $+1/2$ is conserved.

These walls have a variety of interesting properties. Thus, for example, 
if the theory contains fermions, their charge may be half integer. 
In this case the
charges of the wall and the fermion
are exchanged in the scattering process and there is no threshold energy 
for this process. Moreover the fermion and the wall can form a neutral
bound state. For a detailed discussion see Ref.  \cite{kogan}.

Under what circumstances is this wall stable?
The five-dimensional Maxwell term 
$(1/4e_{5}^{2})F_{MN}^{2}$
gives rise to the four-dimensional kinetic term 
$(1/4e^{2})\partial_{\mu}A_{5}\partial_{\mu}A_{5}$ where $e$ is 
the four-dimensional
gauge coupling. It is quite evident that to get a stable domain wall 
an  effective potential  (which does not exist
 at the  classical level) must be generated. 
Such  an effective potential
$U(A_{5}) $
 is indeed induced  at the one-loop level.
The mass operator for the scalar field modes 
 is
\bq
\hat{m} = -i\partial_{5} + A_{5}; \;\;\;\; m_{n} = \frac{n}{R} + A_{5}\, .
\eq
The mass spectrum depends on the value of $A_{5}$  with 
periodicity
$1/R$.
 Due to this dependence there is an effective potential for the
field $A_{5}$ which is given, at one  loop, by
(see \cite{belkogan} for details)
\bq
V(A_{5}) = \frac{1}{(2\pi R)^{4} 16\pi^{2}}\int dt t^{3}
\ln(1-2e^{-t}\cos 2\pi R A_{5} + e^{-2t})\, .
\eq
This potential is periodic in $A_{5}$, with the 
period $1/R$. It   has minima at
 $A_{5} = n/R$.  Using this effective potential it is
not difficult to see that the wall  width $L$ is  of the order of $R/e$
and the  energy density (wall tension)  scales as
 $$ L (1/e^2) (1/RL)^2 = (1/e) (1/R^3).$$
Observe that the wall tension 
 scales as $1/e$ rather than $1/e^2$. This is essentially the  same  
difference as between the $N$ and $N^2$ 
scaling laws in SUSY gluodynamics. Indeed,  $1/N$ 
is an
 effective coupling constant in the string  description of the gauge
theory.  

The same 
$N$ dependence arises also in the non-Abelian case. 
If there are no fields transforming 
in the  fundamental representation,  the  gauge group $G$  is actually
not $SU(N)$ but
$SU(N)/Z_{N}$  and $\pi_{1}(G) = Z_{N}$, so one may have 
\footnote{Let us note that these walls are  nothing but a
$4+1$-dimensional 
generalization of  $Z_{N}$ domain walls  in gluodynamics at finite
temperature \cite{gpyw}. The only difference is that for finite-$T$
theory we must have anti-periodic boundary conditions for fermions,
while in this case  there is no such  requirement.} a $Z_{N}$
domain wall.
The domain walls will be somewhat more
complicated: instead of one phase field there are $N-1$ 
independent
phases (=rank of the group $SU(N)$). 

In the supersymmetric theory the situation is essentially different.
The fermions and the bosons give contributions of the opposite signs
to the effective potential, which, thus,  cancel each other. This is so since 
supersymmetry should be
maintained at any value of $A_{5}$ and the vacuum energy  must be zero.
Therefore, in five-dimensional SUSY gauge theory (which is
essentially a low-energy limit of Witten's theory) there are no stable 
domain walls of the type just discussed.
To get a non-zero  one-loop
 effective potential supersymmetry must be broken explicitly.
This can be achieved either by imposing
different boundary conditions on bosons and fermions, or by adding
non-SUSY
 mass terms by hand, or in any other way --  in any case there {\it will be
 an induced effective potential}.
  In the particular case of $M^4 \times U(1)$ 
relevant  details  can be found in Ref. \cite{belkogan},  and 
the result for the wall tension in the large $N$ 
limit is
\beq
\varepsilon \sim \frac{N}{R^3} \frac{1}{e\sqrt{N}}\, .
\eeq
At large $N$  the proper coupling constant is not $e$ but
 $g = e \sqrt{N}$,  and the domain wall tension  has
 a typical  $D$-brane behavior -- linear in both $N$ and $1/g$,
$$
\varepsilon \sim (N/g) (1/R^3)\, . 
$$

 Thus we see that in the  theory of the type considered in \cite{witten} 
(with explicitly broken  SUSY) 
a  new type of the domain wall, with an abnormal $N$-dependence, emerges.
The $D$-brane interpretation of these walls should also be possible.

\section{Comment on Discrete Anomaly Matching}

The issue of the discrete anomaly matching in SUSY gluodynamics 
was raised in Ref. \cite{CH}, with the conclusion that
the chirally symmetric vacuum suggested in Ref. \cite{kovner2}
does not satisfy the matching condition, at least in its naive
realization. Although this topic is peripheral to the main subject of 
the present work, we would like to dwell briefly on this issue in view of 
general important consequences  which might follow. 

The construction which goes under the name of {\em discrete 
anomaly matching} was suggested in Refs. \cite{1,2,3}, in a 
model-building 
context where it was quite informative  and  useful. Below we will 
show that in supersymmetric  gluodynamics and similar settings
no new physical results can be obtained from the procedure,
besides those Ward identities which are already obtained by a 
different method. These Ward identities neither support nor rule out 
the chirally symmetric Kovner-Shifman vacuum.

Let us introduce the construction in an explicit form, first in a 
somewhat simplified setting of a non-supersymmetric Yang-Mills 
theory, with the intention to revisit SUSY gluodynamics 
later on. Assume we have a $SU(N)$ Yang-Mills theory
with one quark field $\Psi^a$ belonging to the adjoint representation 
of the gauge group. Note that, unlike SUSY gluodynamics,
$\Psi$ is the Dirac field, so it can be coupled to an 
external ``electromagnetic" field $A_\mu$ vectorially, with a very 
small coupling constant $e$,
\beq
\Delta {\cal L} = e\bar\Psi^a \gamma^\mu\Psi^a A_\mu\, .
\label{vecver}
\eeq
The field $A_\mu$ is auxiliary and is needed only for the
purpose of constructing the 't Hooft AVV triangles \cite{THmatch}.
The quark $\Psi^a$ is assumed to be massless.

This theory, classically,  has two conserved currents, the vector 
current $\bar\Psi \gamma^\mu\Psi$ and the axial one, $\bar\Psi 
\gamma^\mu\gamma^5\Psi$. (There are two other conserved currents \cite{Kosh}, but
this is another story.) The vector current is useless from the 
point of view of the 't Hooft matching \cite{THmatch} and we will 
forget about it,
 focusing on the axial $U(1)$ symmetry associated with
the axial current. This symmetry is internally anomalous,
\beq
\partial^\mu \left( \bar\Psi \gamma^\mu\gamma^5\Psi \right) = 
\frac{N}{8\pi^2}
G_{\mu\nu}^a\tilde{G}_{\mu\nu}^a\, ,
\label{inam}
\eeq
so that  the only remnant of this ``non-symmetry" is a discrete 
$Z_{4N}$. The fact of survival of the discrete $Z_{4N}$ symmetry
is most readily seen from the instanton-induced 't Hooft interaction 
\cite{GTH} which in the case at hand includes $4N$ fermion lines. 
The factor $4N$ is related to the coefficient $N$ in the right-hand 
side of Eq. (\ref{inam}), which in turn represents $T(G)$, (one half) of 
the Dynkin index for the adjoint representation. If instead of the 
adjoint quark we were dealing with the quark in a representation 
$R$, then $T(G)\ra T(R)$. 

The survival of a discrete unbroken subgroup of the axial $U(1)$ is 
similar
to what we have in SUSY gluodynamics. 

To construct the 't Hooft triangles that must be matched at the 
fundamental and constituent levels we have to have a continuous 
axial symmetry, rather than a discrete one.  The basic idea
of the discrete matching \cite{1,2,3} is embedding the theory under 
consideration into a larger one, where we do have a continuous axial 
symmetry, which is later spontaneously broken down to $Z_{4N}$. 
The spontaneous breaking should happen in such a way 
that at low energies we 
recover the original theory  plus possible extra decoupled degrees of freedom.  
Such an embedding is easy to achieve by exploiting the phantom 
axion construction \cite{SVZ}.
Let us add to our Yang-Mills theory a quark field $Q$ in the 
fundamental representation of the gauge group. This quark is 
coupled to a (color-singlet) complex scalar field $\Phi$,
\beq
\Delta {\cal L}_\Phi = h \bar{Q}_R Q_L \Phi + \mbox{H.c.} + V(|\Phi 
|)\, ,
\label{embed}
\eeq
where $h$ is a coupling constant, and $V(|\Phi |)$ is a self-interaction 
potential which will eventually ensure the development of a large 
vacuum expectation value of the field $\Phi $,
$$
\langle \Phi \rangle = v \ra \infty\, .
$$
When $v$ becomes very large, the quark $Q$ and the modulus of the 
filed $\Phi$ disappear from the spectrum, leaving only Arg$\Phi$, 
the axion field, as a remnant. The quark $Q$ and the modulus of the 
filed $\Phi$  are auxiliary elements of the construction. 
With the extra fields introduced in this way we have an additional
axial $U(1)$ symmetry
\beq
Q_L \ra Q_L e^{i\beta }\, , \,\,\,  Q_R \ra Q_R e^{- i\beta }\, , \,\,\,
\Phi \ra \Phi e^{- 2 i\beta }\, .
\label{addasymm}
\eeq
This symmetry is internally anomalous too. Out of two internally 
anomalous $U(1)$ symmetries we can readily pick up an 
anomaly-free combination,
\beq
\psi_L \ra \psi_L e^{- i\alpha}\, , \,\,\, 
Q_L \ra Q_L e^{-i 2N \alpha }\, , \,\,\, 
\Phi \ra \Phi e^{i 4N \alpha }\, .
\label{consasymm}
\eeq
The corresponding conserved axial current has the form
\beq
J_\mu = 
\bar\Psi \gamma^\mu\gamma^5\Psi - 2N 
\bar{Q}\gamma^\mu\gamma^5Q +
\mbox{scalar term}\, .
\label{fac}
\eeq
The vacuum expectation value of the field $\Phi$ spontaneously 
breaks
the $U(1)$ symmetry of Eq. (\ref{consasymm}), but since the 
corresponding charge of the $\Phi$ field is $4N$, there is a survivor, 
a $Z_{4N}$ subgroup.

So, we have a theory which, at the fundamental level, has an 
internally anomaly-free axial $U(1)$ containing  a discrete 
subgroup.
At the scale below $v$ only the discrete subgroup survives.
Let us examine now implications of the 't Hooft anomaly matching.
The triangle to be considered is AVV where the A vertex is due to 
the axial
current (\ref{fac}), while the vector vertices are due to 
Eq. (\ref{vecver}). Note that the auxiliary quark $Q$ has no coupling 
to $A_\mu$.

The  AVV triangle appearing at the fundamental level is
\beq
\partial^\mu J_\mu = \frac{N^2 -1 }{8\pi^2}F_{\mu\nu}
\tilde{F}_{\mu\nu}
\label{dfl}
\eeq 
where $F_{\mu\nu}$ is the photon field strength tensor built of the 
auxiliary field $A_\mu$. If the photons are on mass shell, Eq. 
(\ref{dfl}) implies \cite{DZ,THmatch}
the existence of a pole coupled to $J_\mu$,
\beq
\langle J_\mu\rangle = \frac{q_\mu}{q^2}\frac{N^2 -1 
}{8\pi^2}F_{\mu\nu}
\tilde{F}_{\mu\nu}\, .
\label{fundpol}
\eeq
The coefficient in front of $q_\mu / q^2$ has to be matched by the 
contribution of physical massless particles. Some of them may or 
may not occur dynamically, as composite mesons or baryons built 
from $\Psi$'s  in the original Yang-Mills theory under consideration.
More important  is the occurrence of the massless axion field, 
which is coupled to the current $J_\mu$ and, thus, participates
in the matching with necessity. This is a distinctive feature of the 
discrete matching, as opposed to the 't Hooft matching, 
where such  field, totally foreign to the original Yang-Mills 
theory {\em per se}, does not emerge. It is to be stressed that, 
as opposed to the Peccei--Quinn construction \cite{PQ},  
in the present setup the axion is necessarily massless and can not acquire mass
through nonperturbative effects. This is so since it is a Goldstone boson
appearing due to spontaneous breaking of a continuous global symmetry.

The axion field $a$  is not coupled to the $A_\mu$ field because
the auxiliary quark $Q$ does not have this coupling. It is coupled, 
however, to the gluon field, through the standard vertex
\beq
a \, \frac{1}{32\pi^2}\, G_{\mu\nu}^a\tilde{G}_{\mu\nu}^a\, .
\eeq
Since its coupling to the current $J_\mu$ is
\beq
J_\mu^{\rm axion} = - 4N v^2 (\partial_\mu a)
\label{jax}
\eeq 
we conclude that at low  energies, in the effective low-energy theory,  $\langle J_\mu\rangle =
\langle J_\mu\rangle ^{\rm axion}$
plus a possible pole term in $\langle J_\mu\rangle$ due
to  the contribution of massless composites built of $\Psi$,
should they exist. Here 
\beq
\langle J_\mu\rangle ^{\rm axion} = \frac{q_\mu}{q^2} \left(-\frac{N}{8\pi^2} 
\right)\, \langle 0 | G_{\mu\nu}^a\tilde{G}_{\mu\nu}^a| 
\gamma\gamma\rangle\, .
\label{eleth}
\eeq
The momentum $q$ in Eq. (\ref{eleth})
is the momentum flowing in the $G\tilde G$ vertex
(the total momentum of the photon pair). It is assumed that
$q \ra 0$. 

The matching of Eqs. (\ref{fundpol}) and (\ref{eleth})
tells us  that
\beq
\frac{N^2 -1 
}{8\pi^2}F_{\mu\nu}
\tilde{F}_{\mu\nu} +\left(\frac{N}{8\pi^2} \right)\, \langle 0 | 
G_{\mu\nu}^a\tilde{G}_{\mu\nu}^a| \gamma\gamma\rangle
\label{nsvzanom}
\eeq
$$
\mbox{= possible contributions due to massless $\Psi$ composites}.
$$
If there are none, then the expression on the left-hand side vanishes.

Recall that our initial  task was getting information on
the emergence or nonemergence of the massless 
$\Psi$ composites. The entire construction with embedding the
discrete remnant of the anomalous axial $U(1)$ was designed for that 
purpose. We are neither closer nor further now from this goal.
Indeed, one can discard this construction altogether, and just 
consider the internally anomalous current (\ref{inam}). Then, 
combining both the external and internal anomaly, we would get
\beq
\partial^\mu \left( \bar\Psi \gamma^\mu\gamma^5\Psi \right) = 
\frac{N}{8\pi^2}
G_{\mu\nu}^a\tilde{G}_{\mu\nu}^a + \frac{N^2-1}{8\pi^2}
F_{\mu\nu}\tilde{F}_{\mu\nu}\, .
\label{toyanom}
\eeq
Sandwiching both sides of this formula between $\langle 0 |$
and $|\gamma\gamma\rangle$ in the limit $q\ra 0$ we immediately reproduce 
Eq. (\ref{nsvzanom}). The only interesting dynamical question is 
whether the left-hand side of Eq. (\ref{nsvzanom}) vanishes or not. 
At first sight the 
$N$ dependence of two terms in this equation is different, so one is 
tempted to say that they cannot cancel each other. 
A closer look shows, however, that the discrepancy is superficial. 
Indeed, a typical graph for the second term is depicted in Fig. 5.
The gluons are converted into photons through the $\Psi$ loop.
It is not difficult to count that the matrix element shown in Fig. 5 
scales as $N^2$, i.e. in the same way as the first term  in Eq. 
(\ref{nsvzanom}).

\begin{figure}
  \epsfxsize=7cm
  \centerline{\epsfbox{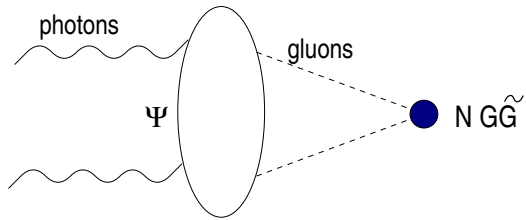}}
  \caption{ The two-photon matrix element of $NG\tilde G$. The 
photons are assumed to be on mass shell.}    
\end{figure}

If $T(R)$ were a free parameter than one could establish
the non-vanishing of Eq. (\ref{nsvzanom}) since the second term is 
proportional to $(T(R))^2$ while the first term to dim$R$.
The choice of $T(R)$ is not free, however, since, on the one hand,
to have a discrete unbroken subgroup $Z_N$ we must work with the 
quarks in representation higher than fundamental, but on the other
hand
the representation can not be too high,
since otherwise we loose asymptotic freedom. For this reason $T(R)$ 
cannot scale 
faster than $N$. These two requirements are contradictory unless 
$T(R)\propto N$.

Thus, Eq. (\ref{nsvzanom})  may or may not vanish, depending on 
whether the two terms cancel each other. As far as the 
$N$ dependence is concerned, they are perfectly fit to 
cancel. In the absence of massless
composites they {\em would be forced} to cancel.
This is nothing but the NSVZ low-energy theorem for 
the two-photon coupling to $G\tilde G$ \cite{NSVZ}. 
  
Instead of the auxiliary photon $A_\mu$ we could have considered 
the coupling to gravitons, i.e. the $U(1)$ current in the gravitational 
background. Then the issue would reduce to a formula connecting 
$(N^2 - 1)R\tilde R$ to a two-graviton matrix element of $NG\tilde 
G$. Again, the so called discrete matching would have nothing to say
whether or not the two terms combine to cancel each other (in the 
first case  there are no
massless $\Psi$ composites while in the second they would have to 
be present to match the anomaly). 

Now, we can readily adapt our consideration to SUSY gluodynamics. 
Again, we could have built a ``tower of discrete anomaly matching" 
by embedding the theory in a larger one where an internally
nonanomalous axial $U(1)$ current exists, with the subsequent 
spontaneous breaking of this $U(1)$ down to $Z_{2N}$, the actual 
symmetry of SUSY gluodynamics. As we have just demonstrated,
this procedure is redundant. It would yield no more constrains or 
information 
compared to what one gets considering the axial current of gluinos 
from the very beginning. The gluino is described by the Majorana 
field, so we cannot couple it to the auxiliary photon (vector current).  
However, the anomaly in the gravitational background remains an 
open possibility. The relation to be analyzed is
\beq
\partial_\mu \left( 
\bar\lambda\gamma^\mu\gamma^5\lambda\right)
=\frac{N}{16\pi^2}\langle 0 | 
G_{\mu\nu}^a\tilde{G}_{\mu\nu}^a|
\mbox{2 gravitons}\rangle  + \frac{N^2-1}{8\pi^2}
\, C\, R\tilde R\, ,
\label{ksanom}
\eeq
where $C$ is a known constant. The question to be answered is: are 
there massless composites built from gluons/gluinos?
In the standard chirally asymmetric 
phase we expect none, while in the chirally symmetric vacuum of 
Kovner and Shifman a set of massless composites must exist.

We see that, if at all, the massless composites of the Kovner-Shifman 
solution facilitate the anomaly matching.
Indeed, in the chirally asymmetric vacua the exact cancellation of 
two terms in Eq. (\ref{ksanom}) must take place, while in the chirally 
symmetric one
this cancellation can be partial. The missing part will  then be filled 
by the contribution of massless composites. Regardless, the $N$
dependence of both terms in  Eq. (\ref{ksanom})
is the same, and no constraints on the chirally symmetric solution 
\cite{kovner2}  follow\footnote{We note that the existence of 
the chirally symmetric
phase was questioned recently on different grounds in \cite{SZ}. 
It was claimed that 
such a chirally symmetric phase would be necessarily 
superconformally invariant and, therefore, have
more symmetries than the Lagrangian of the original theory.
Unfortunately this argument is not substantive. 
First, the fact of superconformal invariance of the chirally symmetric phase
was not established in \cite{SZ}. It is perfectly conceivable that the 
correlators in this phase depend logarithmically on $\Lambda_{QCD}$. 
Second, even if the superconformal invariance is there, 
this is not forbidden by 
general principles of quantum field theory. For instance, in the 
realm of models of 
critical phenomena, the phenomenon of symmetry enhancement at the 
infrared fixed point
is well known and not at all rare.}.

To summarize,  the addition of extra 
matter which promotes the discrete $Z_{2N}$
symmetry into the continuous one necessarily leads to the appearance of a 
massless axion. This axion is indeed practically decoupled from the dynamics
of the rest of the low energy sector when the symmetry 
breaking scale $v^2$ is large.
However, at the same time it couples strongly to the conserved global current
by virtue of Eq. (\ref{jax}). As a result, the contribution of the 
axion to the 
anomaly matching is finite and independent of the scale $v$. This 
contribution is
the first term on the right hand side of Eqs. (\ref{toyanom}) 
and (\ref{ksanom}). 
The discrete anomaly matching conditions, therefore, do not 
pose any restrictions on the 
spectrum of massless composite fermions but, rather, just 
determine the contribution of the
axion which is not a physical quantity in the original theory.
In this language the ``modulo $N$" matching of Ref. \cite{CH} is the 
statement that
the contribution of the axion to an anomaly triangle has to be an integer 
multiple \footnote{We use the phrase ``integer multiple of $N$" 
in a somewhat loose 
sense. This contribution depends on what particular anomaly 
triangle one considers
and on some other details, e.g. the existence of massive Majorana fermions in
the spectrum. These details 
are unimportant for us here. For a thorough discussion see 
Refs. \cite{2}, \cite{3} and \cite{CH}.} of $N$.
This statement is true in a simple case when the fundamental fermions 
acquire mass only 
due to a  Yukawa coupling to the scalar whose vacuum expectation value
 breaks the global 
$U(1)$ symmetry down to
the discrete subgroup in question.
It is, however, not provable as 
a general result
and there is no {\it a priori} reason to believe that it holds in 
strongly interacting 
theories with confining dynamics, like SUSY gluodynamics.

For example, let us consider the very same 
 toy model (\ref{embed}), but this time, instead of the vector
current (\ref{vecver}), let us analyze the   
triangle with two vector currents $\bar Q\gamma_\mu Q$
(and the same axial current as above).
For convenience one can couple $\bar Q\gamma_\mu Q$ to another auxiliary vector field
${\cal A}_\mu$, which is distinct from the field $A$ introduced above.
(${\cal A}$ does not couple to $\Psi$.) The gauge strength tensor built from 
${\cal A}_\mu$ will be denoted by ${\cal F}_{\mu\nu}$.
 Now we can match
the anomaly between the high energy scale $M\gg v$ where the
extra quarks $Q$ are massless and the intermediate scale $m$
($v\gg m \gg \Lambda$). At this scale the quarks $Q$ do not appear in the spectrum
anymore. Note that the truly dynamical quarks $\Psi$ and their scale
$\Lambda$ are irrelevant in this problem.

In this matching the contribution of the massless axion at the scale
$m$ must be  equal to the contribution of the fermions $Q$ at the scale $M$,
since the only nontrivial dynamics that happens at $v$ is the
spontaneous symmetry breaking due to the Higgs field $\Phi$. Since $Q$
couple directly to $\Phi$ their axial 
charges are necessarily integer multiples of $2N$, Eq. (\ref{consasymm}).
The contribution of $Q$ to the  anomalous triangle  at $M$ is 
$$
- 4N^2 \frac{1}{16\pi^2} {\cal F}_{\mu\nu}{\tilde{\cal F}}_{\mu\nu}\, ;
$$
the coefficient is 
multiple integer of $4N$. Thus, it is indeed true that the contribution
of the axion to this anomaly  at the intermediate scale $m$ is an
integer multiple of $4N$. 
This is of course trivially so, since the axion coupling 
$$
a \frac{N}{16\pi^2} {\cal F}_{\mu\nu}{\tilde{\cal F}}_{\mu\nu}
$$
in conjunction with Eq. (\ref{jax}) automatically guarantee
the required proportionality.

We are  interested, however,  in matching the
anomaly between $m$ and a still lower scale $\mu<\Lambda$.
We would have the statement about ``modulo N'' matching here if we
knew that the contribution of the axion changes by an integer multiple
of $4N$ when crossing the scale $\Lambda$. This we  can
not know, however,  because of the strong nontrivial interaction at $\Lambda$.
The best we can do is to express this contribution in terms of a
matrix element of $G\tilde G$, see Eq.(\ref{nsvzanom}).
The fact that the axion couples extremely 
weakly to the gauge field $G$ does not
help here, since it is a finite ``renormalization'' of this weak
coupling due the strong interactions of the gauge field and $\Psi$  that
determines the contribution of the axion to the anomaly.

\section{Discussion}

In this paper we have analyzed   aspects of the supersymmetric 
walls in SUSY gluodynamics and in a more general context. We have 
argued that the linear scaling of the
wall tension  with $N$ does not  contradict the picture where  the 
wall is a
classical soliton in an  effective Lagrangian describing low-energy
mesons and glueballs. 
We have also provided a simple qualitative explanation of how the 
confining string can end on a domain wall. 

A key part of our analysis is related to the issue what happens
when the low-energy Lagrangian has a glued structure.  The 
Veneziano--Yankielowicz  description of SUSY gluodynamics belongs 
to this class. 

We have tried to answer the question whether the walls in  SUSY 
gluodynamics  are indeed BPS saturated. Our
conclusion is that the knowledge of the effective Lagrangian by itself 
is 
not
sufficient to answer this question if the wall in question crosses 
boundaries of distinct sectors. The effective Lagrangian has a
cusp structure which arises due to adiabatic integration of the heavy
degrees of freedom. The adiabatic approximation breaks down when 
the wall trajectory crosses the cusp.
We have considered a simple
model which illustrates this feature in detail. In this toy model we
were able to calculate the extra term, which gives the cusp
contribution to the wall energy. Unfortunately in  SUSY
gluodynamics  we are unaware  of a well-defined procedure 
which
could be used to obtain the missing term, since the (amended)
Veneziano--Yankielowicz effective Lagrangian was not obtained by 
explicitly
integrating out  heavy fields, but, rather,  from certain  Ward
identities of the theory. It is an interesting question whether these
same Ward identities could also determine the cusp contribution.

We then made some observations of the general nature pertinent to 
the theory of the domain walls in the effective low-energy theories.
Supersymmetric walls possess some unique features. Namely, if we 
find the BPS wall tension (for a wall which does not cross the sector 
boundary) in the effective theory, the very same tension
takes place in the full theory. There are no corrections inversely
proportional to the masses of the heavy fields which were integrated 
out.  

Finally, we worked out the issue of the discrete anomaly matching
in SUSY gluodynamics. This procedure, when appropriately 
implemented, is shown to impose no constraints on the existence of 
the chirally symmetric vacuum state \cite{kovner2}. 

We would like to make a remark on relation of our analysis to
the calculations of \cite{smilga}. The analysis of \cite{smilga} is 
performed in the framework of the effective Taylor - Veneziano - Yankielowicz
(TVY) effective Lagrangian, which in addition to the "glueball" superfield S
contains matter superfields corresponding to additional matter fields
in the SUSY QCD. Due to inclusion of these additional superfields the TVY
Lagrangian does not have a cusp structure and the wall configurations
considered in \cite{smilga} therefore do not cross any cusps.
The numerical analysis of the simplest TVY Lagrangian carried in \cite{smilga}
showed that the BPS saturated wall, although present in the weak coupling
regime - at small value of the Higgs mass - disappears for masses greater
than some critical mass $m_*$. Even more surprisingly at a slightly
greater mass $m_{**}$
even 
the non BPS solutions that connect two adjacent vacua but do not pass through
the chiral point $S=0$ disappear altogether.
Since the VY effective Lagrangian is obtained from the TVY Lagrangian 
in the limit $m\rightarrow\infty$ the authors of \cite{smilga} tentatively
conclude that the only walls that exist in pure SUSY gluodynamics are
the ones that pass through the chirally symmetric vacuum. 
If this is the case it is indeed very surprising, since the energy density
of these walls in the large $N$ limit is $O(N^2)$.
We feel however that it is premature to draw such definite conclusions from
the existing calculations for two main reasons. First, the calculations
have only been performed at $N=2,3$. It is not clear whether
the 
critical mass remains finite 
at large $N$ if the kinetic term in the TVY Lagrangian
is taken to reproduce the correct large $N$ scaling of meson masses.
Secondly, even though the introduction of the matter fields eliminates
the cusps in the effective potential we believe that in the SUSY 
gluodynamics limit this effect must be unphysical. The heavy degrees of freedom
that should be responsible for smoothening of the effective
potential in SUSY gluodynamics should be
glueballs which are heavier than the order parameter field $S$ but
still much lighter than the matter fields, which become infinitely
heavy and should completely
decouple in this limit. To our mind 
it is therefore questionable that the TVY effective
Lagrangian reflects correctly the  physics of heavy modes
at large matter fields masses, where it ceases to be the effective
Lagrangian in the Wilsonian sense.
We think therefore that the existence or nonexistence of BPS
saturated walls in SUSY gluodynamics remains an open question. 
 We also would like to add that after
  this work has been finished we learned about the preprint
\cite{gomez} where the  vacuum structure and domain walls in SUSY gluodynamics
 were studied using D-brane approach.

\vspace{0.3cm}

{\bf Acknowledgments}: \hspace{0.2cm} 

The authors are grateful to G. Dvali for pointing out a sloppy
treatment of one of the models in Sect. 2.1.
We also thank A. Smilga for interesting correspondence regarding
relation of this work and \cite{smilga}.

This work was supported in part by DOE under the grant number
DE-FG02-94ER40823.
A.K. is supported by PPARC advanced fellowship.

\end{document}